\documentclass[universe,article,accept,oneauthor,pdftex]{Definitions/mdpi}

\firstpage{1}
\pubvolume{1}
\issuenum{1}
\articlenumber{0}
\pubyear{2021}
\copyrightyear{2021}
\externaleditor{Academic Editor: Włodzimierz Piechocki}
\datereceived{1 April 2021}
\dateaccepted{23 April 2021}
\datepublished{}
\hreflink{https://doi.org/}

\def\ka{\kappa}

\def\pa{\partial}

\def\al{\alpha}

\def\ka{\kappa}
\def\ii{\textrm i}
\def\ee{\textrm e}
\def\e{\textrm e}

\def\setR{\mathbb{R}}

\newcommand{\be}{\begin{equation}}
\newcommand{\en}{\end{equation}}
\newcommand{\bi}{\begin{itemize}}
\newcommand{\ei}{\end{itemize}}

\newcommand{\dd}{\mathrm{d}}

\Title{The de Broglie--Bohm Quantum Theory and Its Application to Quantum Cosmology}%mdpi: please unify the foramt of de in ``de Broglie--Bohm'' for the whole section titles

\TitleCitation{The de Broglie--Bohm Quantum Theory and Its Application to Quantum Cosmology}

\Author{Nelson Pinto-Neto %mdpi: Please carefully check the accuracy of name and affiliation. Changes will not be possible after proofreading.
\orcidA{}}

\AuthorNames{Nelson Pinto-Neto}

\AuthorCitation{Pinto-Neto, N.}

\address[1]{Centro Brasileiro de Pesquisas F\'{\i}sicas, Rua Dr. Xavier Sigaud, 150, Urca, Rio de Janeiro CEP-22290-180, RJ, Brazil; nelsonpn@cbpf.br; Tel.: +55-21-21417381}%mdpi: we merged the original Current address, please confirm

\abstract{We review the de Broglie--Bohm quantum theory. It is an alternative description
of quantum phenomena in accordance with all the quantum experiments already performed.
Essentially, it is a dynamical theory
about objectively real trajectories in the configuration space of the physical system
under investigation. Hence, it is not
necessarily probabilistic, and it dispenses with the collapse postulate,
making it suitable to be applied to cosmology. The emerging cosmological models are
usually free of singularities, with a bounce connecting a contracting era
with an expanding phase, which we are now observing. A theory of cosmological
perturbations can also be constructed under this framework, which can be
successfully confronted with current observations, and can complement
inflation or even be an alternative to it.}


\begin{document}

\section{Introduction}

In the dawn of modern cosmology, one of the reasons Einstein first rejected the Friedmann solution was because it contains an initial singularity. When he was forced to accept it, he claimed that the initial singularity in the Friedmann universe was General Relativity (GR) pointing us to its own limits of applicability. In fact, GR is generally plagued with singularities~\cite{fried1,fried2,fried3}.
New physics must take place near the singularity, where the curvature of space-time and the energy-density of the matter fields attain immensely high values. This new physics is not evident, because the singularities appear under very general reasonable assumptions.%mdpi: we removed extra bold, please confirm

There are classical extensions to GR that can be proposed, such as non-minimal couplings between matter gravitational degrees of freedom, the addition of curvature squared terms in the gravitational Lagrangian, and  %Please check intended meaning is retained
the presence of exotic matter fields, { many of them coming from effective actions taking into account quantum effects that can eliminate the cosmological singularity; see~\cite{bounce-classical1,bounce-classical2,bounce-classical3,bounce-classical4,bounce-classical5,bounce-classical6,bounce-classical7,
bounce-classical8,bounce-classical9,bounce-classical10,bounce-classical11,bounce-classical12,bounce-classical13,reviews1,reviews2} for some reviews.} Another route, in analogy with the procedure adopted to treat the singularities present in the classical description of matter (the instability of the classical model of the atom, the divergence of the electromagnetic field near the electron), is to expect that a proper { and full} quantization of GR could eliminate these singularities. However, a consensual theory of quantum gravity is not yet available, with many proposals still under construction~\cite{qg1,qg2,qg3,qg4}. Nevertheless, in the case of cosmology, as observations indicate that the Universe was nearly homogeneous when it was very hot and dense, with small inhomogeneous perturbations around this very symmetric state~\cite{CMB,nucleo}, one can design an effective quantum theory for cosmology, where the complete configuration space of GR and matter fields, called superspace, is reduced to a subset containing only the homogeneous and perturbation degrees of freedom, called midi-superspace, where the technical problems surrounding quantum gravity are dramatically simpler. This line of investigation is called quantum cosmology~\cite{qc1,qc2,qc3,qc4,qc5,qc6,qc7,qc8,qc9,qc10,qc11,qc12,qc13,qc14}. The rigorous mathematical justification for this reduction is not yet known,  %Please check intended meaning is retained
but it is expected that, if one is not very close to the Planck length, this approach can not only furnish the right corrections to the classical cosmological models in the extreme physical situations around the singularity, but it can also teach us the sort of properties a complete theory of quantum gravity might have. Some relevant questions are: What is a singularity in quantum space-time? Does a classical singularity survive quantization? How can the classical limit be reached?
What is the meaning of time in quantum space-time?

Nevertheless, beyond all the problems surrounding quantum gravity, there is an extra fundamental question concerning the application of quantum theory to cosmology. As we know, in the usual Copenhagen interpretation~\cite{bohr,hei,von}, the wave function gives the probability density amplitude for an external observer to measure one of the eigenvalues of a Hermitian operator describing an observable of a physical system in state $|\Psi\rangle$. In the measuring process, the system must interact with a measuring apparatus. In the quantum description of the whole process, the total wave function describing the system and apparatus bifurcates into many branches, each one containing one of the possible results of the measurement. However, at the end of the measurement process, just one value is obtained; hence, the total wave function must collapse in one of the branches. This a non-unitary non-linear process that cannot be described by the unitary quantum evolution. The intervention of the classical observer imposes a break on the quantum description, bringing to actual existence the many potentialities the quantum state describes. Of course, one cannot apply this picture to the Universe as whole, as, by the definition of Universe, there is nothing external to it that can bring to actual existence all the potentialities described in a quantum state of the Universe. In this scenario, quantum cosmology does not make any sense; it cannot describe the objective reality we experience in the world; it is an empty theory. One should then abdicate to apply quantum theory to cosmology, in order to use it to solve the classical cosmological problems. This is a good example that corroborates an important criticism of Einstein's concerning quantum theory in the Copenhagen framework~\cite{einstein}: `Contemporary quantum theory … constitutes an optimum formulation of [certain] connections … [but] offers no useful point of departure for future developments.'.

Fortunately, there are alternative quantum theories. One can cite the
Many Worlds Interpretation (MWI)~\cite{eve}, where the wave function does not collapse, and all potentialities take place in each branch, but the branches are not aware of each other, or the Spontaneous Collapse approach, where the unitary Schr\"odinger evolution is supplemented with a non-linear evolution where the collapse of the wave function takes place physically~\cite{rim,pen}, among others. In both approaches, there is no need for an external agent to turn the quantum potentialities into actual facts. { These alternative quantum theories have been used in quantum cosmology; for some examples, see~\cite{qc3,mwiqc2,rimqc1,rimqc2,chqc1,chqc2}.}

The framework we will use here is de Broglie--Bohm quantum theory~\cite{bohm,hol,duerr}. In this approach, the point in configuration
space describing the degrees of freedom of the physical system and the measuring apparatus is supposed to exist,
independently of any observations. This point refers to either particle positions and/or field configurations. In the splitting of the total wave function, the point in configuration space will enter
into one of the branches, depending on the initial positions
before the measurement interaction, which are unknown. The other branches will be empty, and it can be shown~\cite{bohm,hol,duerr} that the empty waves can never interact with the actual degrees of freedom describing the physical system, measuring apparatus, or any other external agent.
Hence, no
observer can be aware of these empty branches. We thus have
an effective but not real collapse, as the empty waves continue to exist, but now, contrary to MWI, with no multiplication of actual worlds. There is only one actual world and a profusion of empty innocuous empty waves.
As in the MWI and Spontaneous Collapse approaches, the presence of an external agent is not necessary for understanding quantum measurements, and the quantum dynamics are always valid. In these frameworks, quantum theory can be viewed as the fundamental theory of Nature, applicable to all physical systems, including the Universe itself, from which classical mechanics is a by-product, under certain physical conditions. Hence, in these formulations of quantum theory, quantum cosmology makes sense, and it can be~studied.

In this paper, I will summarize the results that were obtained with the application of the dBB quantum theory to quantum cosmology. In fact, the assumption that particle positions and/or field configurations are actual makes quantum cosmology conceptually simpler. The concepts of quantum singularities, how they can be removed, and the classical limit can be easily obtained. The notion of time emerges naturally from timeless quantum dynamics, and the Schr\"{o}dinger equation for quantum inhomogeneous perturbations in quantum homogeneous backgrounds is dramatically simplified under the dBB assumptions, allowing the calculation of many cosmological observable quantities. Finally, a sound interpretation of the wave function of the Universe emerges in this framework; see \mbox{Section \ref{sec3}}. { In technical terms, however, the dBB quantum theory adds further computational difficulties, as it may require the calculation of the quantum trajectories of the quantum degrees of freedom, which can be extremely hard in general. However, in the framework of cosmology, the necessary computations to be performed are not difficult, and they help the construction of simple extensions of the quantum equations for quantum cosmological perturbations, as we will see in this paper.}

The review will be divided as follows: In Section \ref{sec2}, I will summarize the main features of the dBB quantum theory. In Section \ref{sec3}, I will apply it to quantum cosmology, considering first the homogeneous background, in order to discuss the singularity problem, and then the inhomogeneous perturbations, where a notion of time emerges. In Section \ref{sec4}, I will present some important results concerning the evolution of quantum cosmological perturbations in quantum backgrounds without singularities, and their confrontations with observations and inflation. In Section \ref{sec5}, I will show that the dBB approach explains, in a quite simple way, an old controversy: the quantum-to-classical transition of the quantum inhomogeneous cosmological perturbations that evolved to form all the structures in the Universe, which are, of course, classical. I finish in Section \ref{sec6} with a discussion and conclusions.

\section{The de Broglie--Bohm Quantum Theory}\label{sec2}

A good way to motivate the construction of the dBB quantum theory in the framework of non-relativistic particles is through the words of John Stewart Bell~\cite{bell}:

`The kinematics of the world, in this orthodox picture, is given by a
wave function for the quantum part, and classical variables - variables which {\em have} values - for the classical part:
$(\Psi(t,q ...), X(t) ...)$. The $X$’s are somehow macroscopic. This is not
spelled out very explicitly. The dynamics is not very precisely
formulated either. It includes a Schr\"{o}dinger equation for the
quantum part, and some sort of classical mechanics for the
classical part, and `collapse’ recipes for their interaction.
It seems to me that the only hope of precision with the dual $(\Psi,x)$
kinematics is to omit completely the shifty split, and let both $\Psi$ and $x$
refer to the world as a whole. Then the $x$’s must not be confined to
some vague macroscopic scale, but must extend to all scales’.

Following Bell's proposal, particle positions must also be considered in order to completely determine the state of a quantum system. Then, besides the Schr\"{o}dinger equation for $\Psi$, one must postulate an equation for $x$. From the Schr\"{o}dinger equation for a single non-relativistic particle in the coordinate representation,
\begin{equation}
\label{bsc}
i \hbar \frac{\partial \Psi (x,t)}{\partial t} = \biggl[-\frac{\hbar ^2}{2m} \nabla ^2 +
V(x)\biggr] \Psi (x,t),
\end{equation}
where $V(x)$ is the classical potential, and writing $\Psi = R \exp (iS/\hbar)$, one obtains the following two real equations:
\begin{equation}
\label{bqp}
\frac{\partial S}{\partial t}+\frac{(\nabla S)^2}{2m} + V
-\frac{\hbar ^2}{2m}\frac{\nabla ^2 R}{R} = 0,
\end{equation}
\begin{equation}
\label{bpr}
\frac{\partial R^2}{\partial t}+\nabla .\left(R^2 \frac{\nabla S}{m}\right) = 0.
\end{equation}

Equation \eqref{bqp} looks like a Hamilton--Jacobi equation for $S$, with an extra term at the end. On the other hand, Equation~\eqref{bpr} can be viewed as the continuity equation for a density distribution of an ensemble of particles given by $R^2$, with $\nabla S/m$ being the velocity field of this ensemble (the quantum current divided by $R^2$). Hence, both equations suggest the following postulates~\cite{bohm,hol,duerr}:

(i) Quantum particles follow objectively real trajectories $x(t)$. They must satisfy the so-called guidance equation:

\begin{equation}
p=m\dot{x}=\nabla S,
\label{guidance1}
\end{equation}
or, equivalently, as first proposed by de Broglie,

\begin{equation}
v(x(t),t)\equiv \dot{x}=\frac{J}{R^2},
\label{guidance2}
\end{equation}
where $v$ is the velocity field, and $J$ is the usual quantum current $J={\rm{Im}}(\hbar\Psi^*\nabla\Psi/m)$.

(ii) The particles are never separated from a quantum field
$\Psi$, which acts on them through Equation~\eqref{guidance1} and satisfies the Schr\"{o}dinger
Equation~(\ref{bsc}).

These are two first-order equations in time, which demand knowledge of initial positions $x_0$ and initial quantum field configurations $\Psi (x,0)$ to be solved uniquely. Initial field configurations can usually be obtained through the preparation of the quantum system, by measuring a complete set of observables in the system. The initial position of the particle, instead, cannot be obtained without disturbing the quantum system. Hence, one cannot exactly know the position of the particle; it is the hidden variable of the theory.

{ Note that solving Equation~\eqref{guidance1} can be very difficult, especially in a many-particle system or field theory. However, as we will see below, both Equations (\ref{bsc}) and \eqref{guidance1} lead to the same probabilistic results as in Copenhagen quantum theory; hence, one can use the usual mathematical techniques to derive these probabilities. Furthermore, in quantum cosmology, what one usually needs is the quantum trajectories (also called the Bohmian trajectories) of the background geometry only, which are not difficult to calculate. Even in the case of quantum inhomogeneous perturbations, as they are supposed to originate from an adiabatic vacuum quantum state, their quantum trajectories are also simple to calculate, as we will see in Section \ref{sec5}. Hence, obtaining solutions from Equation~\eqref{guidance1} will not be problematic in the physical situations with which we are dealing in this paper.}

Because Equations~(\ref{bqp}) and \eqref{guidance1} can indeed be viewed as a Hamilton--Jacobi equation
for the particle, which suffers from the influence of a new quantum potential, besides the classical potential $V$, given by
\begin{equation}
\label{qp}
Q \equiv -\frac{\hbar ^2}{2m}\frac{\nabla ^2 R}{R}.
\end{equation}

\noindent from  %Please check intended meaning is retained
both Equations~(\ref{bqp}) and \eqref{guidance1}, one can obtain the equation of motion:
\begin{equation}
\label{beqm}
m \frac{d^2 x}{d t^2} = -\nabla V - \nabla Q.
\end{equation}

In a statistical ensemble of particles in the quantum
state $\Psi$, if the probability density for the unknown initial
position is given by $P(x_0)=R^2(x=x_0,t=t_0)$, Equation~(\ref{bpr})
guarantees that $R^2(x,t)$ will give the distribution of positions
at any time, and all the statistical predictions of quantum mechanics are recovered. The distribution $R^2$ is called a typical
distribution~\cite{duerr}. Note, however, that, in a fundamental theory describing the dynamics of quantum particles, there is no logical connection between
the distribution of the unknown initial positions with $R^2$. Nevertheless,
whenever $P\neq R^2$, Equations~(\ref{bpr})
and \eqref{guidance1} make $P$ rapidly relax to $R^2$, at least at a coarse
grained level, in many circumstances. This is an analog of the $H$-theorem of statistical mechanics
applied to quantum mechanics; see~\cite{valentini1} for details.
Hence, it seems that the dBB dynamics push physical systems to the typical distribution $P=R^2$, also called the quantum equilibrium distribution.
Note that, if one can find physical systems that
have not relaxed to $P=R^2$, then their statistical predictions will
not agree with conventional quantum mechanics, and the dBB theory could be tested.
The possibility of the  existence of such systems, such as relic gravitational
waves, is now under investigation~\cite{valentini2}.
In conclusion, probabilities are not fundamental
in this theory, and if the tendency to quantum equilibrium is really general, then it may not be necessary to postulate the Born rule, as it could be obtained
through the dynamics themselves.

Let us make some final comments:

(a) The $\Psi$ field guides the particle motion through Equation~\eqref{guidance1}, whoseh effects can be encoded in the quantum potential $Q$. In the case of a many-particle system, when $\Psi$ is entangled, $Q$ can be highly non-local. This is
very important, because the Bell's inequalities~\cite{bell-paper}, together
with Aspect's experiments~\cite{aspect}, show that, in general, a
quantum theory must be non-local, which is the case of
the dBB quantum theory. Additionally, while solving the Schr\"{o}dinger equation, boundary conditions are usually imposed on it, as in the two-slit experiment. Hence, the $\Psi$ field contains this information, which is transmitted to the particle motion through Equation~\eqref{guidance2}. In other words, although there is no classical potential along the route of the quantum particle towards the screen, the quantum potential is not null; it encodes the information contained in $\Psi$, leading to a Bohmian trajectory that is quite complicated. When taken together, the different Bohmian trajectories of an ensemble of quantum particles with an initial position distribution given by $R^2(x,t=0)$ yield the interference pattern typical of the two-slit experiment. Contextuality is also present in the dBB quantum theory.

(b) The classical limit is very simple: we have only to find
the conditions for having $Q\approx0$ when compared with the classical kinetic and
potential energy terms.

(c) Note that, although assuming the ontology of the position of particles in space through the new proposed guidance relation (\ref{guidance1}),
the dBB theory has, at least, the same number of postulates as the Copenhagen interpretation, as long as it dispenses with the collapse postulate. If the Born rule can also be justified under this framework, which is still under debate~\cite{duerr,valentini1,valentini}, then the dBB theory is logically simpler than the Copenhagen interpretation, as it would have one postulate less.

A detailed analysis of the dBB theory in the context of quantum field theory can be seen in Refs.~\cite{hol,oqed,goldPRL,gold2,struyveN}.%mdpi: refs 59 and 60 are missing, please modify
 Generally, it is assumed that field configurations are actual. The probabilistic predictions are in accordance with Poincar\'{e} invariance, but the hidden Bohmian evolution of the fields may violate this symmetry.

Let us end this section in the same way it began, with some words of John Bell~\cite{bell}:

`In 1952 I saw the impossible done. It was in papers by David Bohm.
… the subjectivity of the orthodox version, the
necessary reference to the ‘observer,’ could be eliminated. . . . But why
then had Born not told me of this ‘pilot wave’? If only to point out
what was wrong with it? Why did von Neumann not consider it? . . .
Why is the pilot wave picture ignored in text books? Should it not be
taught, not as the only way, but as an antidote to the prevailing complacency?
To show us that vagueness, subjectivity, and indeterminism,
are not forced on us by experimental facts, but by deliberate theoretical
choice?’

\section{The de Broglie--Bohm Theory Applied to Quantum Cosmology: Background and~Perturbations}\label{sec3}

The structure of the Hamiltonian describing the gravitational field and all other non-gravitational degrees of freedom in the Universe, reads
\vspace{-6pt}

\begin{eqnarray}
\label{ham-gen}
H &=&\int \dd^3x \{N(x){\cal{H}}_0[h(x),\pi_h(x),\varphi(x),\pi_{\varphi}(x)]
\nonumber \\ &+&
N^i(x){\cal{H}}_i[h(x),\pi_h(x),\varphi(x),\pi_{\varphi}(x)]\},
\end{eqnarray}
where $N$ and $N^i$ are Lagrange multipliers, the so called lapse and shift functions; $h$ are the gravitational (geometric) degrees of freedom, usually, the space metric of the space-like hypersurfaces; $\varphi$ represents all non-gravitational degrees of freedom; and $\pi_h,\pi_{\varphi}$ are their respective conjugate momenta. The quantities ${\cal{H}}_0$ and ${\cal{H}}_i$ are the super Hamiltonian and super momentum constraints, originated from the invariance of the full theory under general time and space coordinate transformations. They are constrained to vanish, ${\cal{H}}_0 \approx 0$, ${\cal{H}}_i \approx 0$, where the symbol $\approx$ means `weak equality’, in the sense that they are zero, but the Poisson brackets between them and other canonical variables may not be zero.

The Hamilton equations in terms of the Poisson brackets arise as usual:
\begin{eqnarray}
\label{ham-eq}
\dot{h}(x) &=& \{h(x),H\} , \;\;\;\;\; {\dot{\pi}}_h(x) = \{\pi_h(x),H\} \nonumber \\
\dot{\varphi}(x) &=& \{\varphi (x),H\} , \;\;\;\;\; {\dot{\pi}}_{\varphi}(x) = \{\pi_{\varphi} (x),H\},
\end{eqnarray}
yielding the field evolutions in terms of an arbitrary coordinate time $t$. Note that {the Hamiltonian $H$ is null} due to the constraint equations, a feature of any theory that is invariant under time reparametrizations. {All the constraints are first class: the Poisson brackets among themselves are null in the region of phase space where they are satisfied.}

Following the Dirac quantization procedure for constrained systems, the first class constraints, when turned into operators acting on a Hilbert space, must annihilate the wave functional of the Universe $\Psi$ (expressed in the field representation):

\begin{equation}
\label{WDW}
{\cal{H}}_0[\hat{h}(x),\hat{\pi}_h(x),\hat{\varphi}(x),\hat{\pi}_{\varphi}(x)] \Psi[h(x),\varphi(x)]=0,
\end{equation}
and
\begin{equation}
\label{super-momentum}
{\cal{H}}_i[\hat{h}(x),\hat{\pi}_h(x),\hat{\varphi}(x),\hat{\pi}_{\varphi}(x)] \Psi[h(x),\varphi(x)]=0.
\end{equation}
The Schr\"{o}dinger equation $i\partial\Psi/\partial t = \hat{H}\Psi$ only tells us that the wave functional does not explicitly depend on time, as long as
$\hat{H}\Psi=0$ due to the constraint equations (\ref{WDW},\ref{super-momentum}). We are now using natural units $\hbar=c=1$.

Equation \eqref{super-momentum} just implies that $\Psi$ is invariant under space coordinate transformations of the fields. Equation \eqref{WDW} is the so-called Wheeler--DeWitt equation~\cite{WDW}.

These quantum equations render the interpretation of $\Psi$ quite obscure. First, as we have seen, time has disappeared. It is believed that it is hidden in the Wheeler--DeWitt equation, in which one field degree of freedom will play the role of a physical clock. However, apart from some exceptions, as we will see, this clock variable is not transparent. In general, the Wheeler--DeWitt equation has a Klein--Gordon structure~\cite{WDW}, which makes it difficult to assign a probabilistic interpretation for $\Psi$, as is well known. Some investigations have tried to put the Wheeler--DeWitt equation into a Schr\"odinger form, but, when possible, it was achieved only in an implicit form; see~\cite{kuchar} for a discussion on these issues.

However, when one uses the dBB quantum theory, a quite reasonable interpretation of the wave functional of the Universe emerges. As we have seen, in the dBB theory, one also imposes the guidance relations to the actual field configurations, which are supposed to be objectively real. Looking at Equation~\eqref{guidance1}, one can formally write

\begin{equation}
\label{guidance-fields}
\pi_h(x) = \frac{\delta S[h,\varphi]}{\delta h(x)}, \;\;\;\;\;
\pi_{\varphi}(x) = \frac{\delta S[h,\varphi]}{\delta \varphi(x)},
\end{equation}
where $\pi_{h}(x),\pi_{\varphi}(x)$ are the canonical momenta of $h,\varphi$, which can be expressed in terms of the time derivatives of $h,\varphi$, as usual. For instance, in GR, the canonical momenta conjugate to $3$-metric $h^{ij}$ reads
\begin{equation}
\label{37}
\pi _{ij} = \frac{\delta L_{GR}}{\delta ({\dot{h}}^{ij})} =
- h^{1/2}(K_{ij}-h_{ij}K),
\end{equation}
where $K_{ij}$ is the extrinsic curvature given by
\begin{equation}
\label{33}
K_{ij} = \frac{1}{2N}(2D_{(i}N_{j)}-{\dot{h}}_{ij}) ,
\end{equation}
and $D_i$ is the three-dimensional covariant derivative.

The quantity $S[h,\varphi]$ is the phase of the wave functional of the Universe $\Psi[h,\varphi]$. Hence, Equation~\eqref{guidance-fields} yields the evolution of all the fields in terms of coordinate time $t$ once one knows $\Psi$. This induces the proposition of a nomological interpretation of the wave functional of the Universe: it yields the laws of Nature, in the same way as a Lagrangian and/or a Hamiltonian do in classical mechanics, {see~\cite{goldNom}.} Consequently, the wave functional of the Universe has nothing to do with probabilities, which is quite sensible, as one is talking about a single system, the Universe. Furthermore, it is not surprising that the wave functional of the Universe does not depend on an explicit external time parameter and that the equation that determines it is not generally suitable for inducing a probability measure. However, it would be helpful, in this way of thinking, to find boundary conditions for Equation~\eqref{WDW} where a particular solution emerges as {the} wave functional of the Universe, from which the dynamics of all fields are obtained. Some proposals are under discussion; see, for instance,~\cite{hawking,vilenkin}.

The next natural question to pose is how do probabilities emerge in this conceptual framework? Of course, they should arise when one considers subsystems contained in the Universe, where probabilities can naturally be defined. Indeed, in the dBB approach, one can use the notion of conditional wave functions in order to describe subsystems. Let us suppose that the Universe contains only two fields, $\varphi_1$ and $\varphi_2$. Hence, the wave functional is given by $\Psi[\varphi_1,\varphi_2]$. Suppose one can calculate the Bohmian trajectory for $\varphi_1 \rightarrow \varphi_1(t)$.
Then, one can define the conditional wave functional $\Psi_c[t,\varphi_2]=\Psi[\varphi_1(t),\varphi_2]$, which gives all the information about the evolution of $\varphi_2$. Under certain conditions, the original equation for $\Psi$ becomes a Schr\"odinger equation for $\Psi_c$.
In this situation, quantum equilibrium arises~\cite{duerr,valentini,valentini2}, and one can understand $|\Psi_c|^2$ as a probability measure for subsystems discriminated only by $\varphi_2$, in accordance with daily quantum mechanics.

The use of the dBB quantum theory was essential for constructing this whole conceptual framework. Let us now apply it to quantum cosmology. In this reduced framework, most of the problems associated with the full theory, which are still unsolved, are simpler to handle. As we mentioned in the Introduction, cosmological observations inform us that the Universe was very homogeneous and isotropic when it was very hot and dense, with small inhomogeneous perturbations over this very symmetric state. Hence, one restricts the configuration space to a restricted domain in which the geometry of space-time is given by

\begin{equation}
\label{perturb}
g_{\mu\nu}(t,{\bf {x}})=\bar{g}_{\mu\nu}(t)+h_{\mu\nu}(t,{\bf {x}}),
\end{equation}
where
\begin{equation}
\label{linha-fried}
{\rm {d}} s^{2}=\bar{g}_{\mu\nu}(t) {\rm {d}} x^{\mu} {\rm {d}} x^{\nu}=N^{2}(t) {\rm {d}} t^2 -
a^{2}(t)\delta_{ij} {\rm {d}} x^{i} {\rm {d}} x^{j},
\end{equation}
and
\begin{eqnarray}
\label{perturb-componentes}
h_{00}(t,{\bf {x}})&=&2N^{2}(t)\phi(t,{\bf {x}}) \nonumber \\
h_{0i}(t,{\bf {x}})&=&-N(t)a(t)B_{,i}(t,{\bf {x}}) \\
h_{ij}(t,{\bf {x}})&=&2a^{2}(t)\psi(t,{\bf {x}})\gamma_{ij}-E_{,ij}({t,\bf {x}})), \nonumber
\end{eqnarray}
where $i$ represents $\partial/\partial x^i$, and all the quantities in Equation~\eqref{perturb-componentes} are assumed to be very small when compared with the background degrees of freedom.

Note that I am assuming flat space-like homogeneous and isotropic hypersurfaces just for simplicity; all the calculations can be generalized for spherical and hyperbolic cases. Additionally, it is a good approximation, as indicated by cosmological observations.

Additionally for simplicity, I will consider just one matter degree of freedom, described by the scalar field

\begin{equation}
\label{fluid-velocity}
\varphi(t,{\bf {x}}) = \bar{\varphi}(t) + \delta\varphi(t,{\bf {x}}),
\end{equation}
where $\bar{\varphi}(t)$ is the background field and $\delta\varphi(t,{\bf {x}})$ its first order perturbation.

One should also consider vector and tensor perturbations. Vector perturbations are usually not relevant in inflation and bounce scenarios; see~\cite{vector} for details. Tensor perturbations, or primordial gravitational waves, are represented by the transverse-traceless spatial tensor $h_{ij}^{\rm TT}(t,{\bf {x}})$. Its treatment is technically similar but much easier than the scalar perturbation case. The main results for them will be presented below. For details, see~\cite{PPNGW,PPNGW2}.

I will take a conservative point of view, in which the gravitational field dynamics are described by GR, and the matter field, by a scalar field minimally coupled to gravity, described by the general Lagrangian density $p(X,\varphi)$, where $X=g^{\mu\nu}\partial_{\mu}\varphi \; \partial_{\nu}\varphi/2$. When $p=X^n$, it describes a perfect fluid with equation of state $p=w\rho$, where $p$ is the pressure, $\rho$ is the energy density, and $w=1/(2n-1)$ is constant. This Lagrangian density can also describe a canonical scalar field $p=X-V(\varphi)$. In all cases, $p$ is the pressure associated with the scalar field; see~\cite{mukh-book}. Hence, the action reads
\begin{equation}
\mathcal{S}= \mathcal{S}_{_\mathrm{GR}} + \mathcal{S}_\mathrm{fluid}
= -\frac{1}{2 l_{P}^2} \int \sqrt{-g} R {\rm {d}}^4 x - \int \sqrt{-g} p(X,\varphi) {\rm {d}}^4 x,
\label{action}
\end{equation}
where $ l_{P}=(8\pi G_N)^{1/2}\equiv\sqrt{6}\kappa$ is the Planck length in natural units.

By inserting Equations~(\ref{perturb})--(\ref{perturb-componentes}) into the action (\ref{action}), one can construct the Hamiltonian of the system up to second order in the perturbation expansion, through the usual Legendre transformations. After solving the super-momentum constraint and performing suitable canonical transformations, without ever using the background equations of motion, the Hamiltonian can be generally written as

\begin{equation}
\label{hfinal-vinculos-escalares-hidro} H=N\left[ H_{(0)} +
H_{(2)}\right] ,
\end{equation}
where $H_{(0)}$ and $H_{(2)}$ are the zeroth and second order Hamiltonians, yielding the background and linear cosmological perturbation dynamics, respectively. Note that their sum is constrained to zero, which is a consequence of the invariance of GR under time reparametrizations. Let us now show in detail the two interesting cases of a perfect fluid and a canonical scalar field.

\subsection{Perfect Fluids}

For perfect fluids, one has $p=X^n$. The calculations of $H_{(0)}$ and $H_{(2)}$ lead to (see~\cite{PPNscalar} for details)

\begin{equation}
\label{h00} H_{(0)}\equiv
-\frac{P_{a}^{2}}{4a}+\frac{P_{T}}{a^{3\omega}},
\end{equation}
and
\begin{equation}
\label{h02} H_{(2)}\equiv \frac{1}{2a^{3}}\int
{\rm {d}}^{3}x\pi^{2}({\bf {x}})+\frac{a\omega}{2} \int {\rm {d}}^{3}x v^{,i}({\bf {x}})v_{,i}({\bf {x}}).
\end{equation}

In the background Hamiltonian, $P_T$ arises from the canonical transformation

\begin{equation}
T= \frac{1}{c(1+w)} \frac{\varphi}{ p_{\varphi}^{w}} , \qquad P_{T}= c p_\varphi^{1+w},
\label{can}
\end{equation}
where $p_{\varphi}$ is the momentum conjugate to $\varphi$ and
$c=1/(w\sqrt{2}^{1+3w}n^{1+w})$ is a constant. It is thus connected to the matter degree of freedom. Note, however, that it appears linearly in the Hamiltonian; hence, it can be understood as being canonically conjugate to a clock time $T$. Indeed, this is physically motivated from the definition of $T$ in terms of $\varphi$ and $p_{\varphi}$, and the fact that $\varphi$ is a velocity field potential, $V_{\mu} = \partial_{\mu} \varphi / (2X)^{1/2}$.

The second order part, $H_{(2)}$, yields the dynamics of the perturbation field, $v({\bf x})$, which emerges as the single perturbation degree of freedom left.

When quantizing the system, as explained above, the operator version of the first class constraints must annihilate the wave functional
$\Psi[T,a,v({\bf x})]$,

\begin{equation}
\label{schroedinger}
(\hat{H}_{(0)} + \hat{H}_{(2)})\Psi=0.
\end{equation}
This is a case where Equation~\eqref{schroedinger}
assumes a Schr\"odinger form, because a natural time $T$ emerges from the degrees of freedom of the fluid
\begin{eqnarray}
\label{schr-fluid}
i\frac{\partial}{\partial T}\Psi &=&\frac{1}{4} \left\{
a^{(3w-1)/2}\frac{\partial}{\partial a} \left[
a^{(3w-1)/2}\frac{\partial}{\partial a}\right]
\right\}\Psi \nonumber \\&-&\biggl[\frac{a^{3w-1}}{2}\int
d^3x\frac{\delta^2}{\delta v^2({\bf x})}-\frac{a^{3w+1}w}{2}\int
d^3x v^{,i}({\bf x})v_{,i}({\bf x})\biggr]\Psi .
\end{eqnarray}

{Note that the operator ${\hat{P}}_a^2$ present in $\hat{H}_{(0)}$ is multiplied by $a^{3w-1}$ in Equation~\eqref{schr-fluid} (which can be understood as a particular case of the DeWitt metric~\cite{WDW}), yielding a factor ordering ambiguity. When ${\hat{P}}_a$ becomes a differential operator, there is one particular factor ordering that turns $ a^{3w-1} {\hat{P}}_a^2$ into a covariant one-dimensional Laplacian; hence, it is covariant under coordinate redefinitions. As it is a one-dimensional Laplacian, and one-dimensional manifolds are flat, there exists a special function of $a$ that plays the role of a Cartesian coordinate, in which this term can be written as a simple second order derivative. It reads}

\be
\label{transf}
\chi=\frac{2}{3} (1-\omega)^{-1} a^{3(1-\omega)/2}.
\en

{I chose this particular factor ordering when writing Equation~\eqref{schr-fluid}, and I will use the variable $\chi$ in Section \ref{sec4} in order to solve Equation~\eqref{schr-fluid}.}

I will assume that the background is not entangled with the perturbations
\begin{equation}
\label{ansatz1} \Psi[a,T,v({\bf x})]=\Psi_{(0)}(a,T)\Psi_{(2)}[T,v({\bf x})].
\end{equation}

As a consequence, Equation~\eqref{schr-fluid} leads to the equation
\begin{eqnarray}
\label{scrhoedinger-separado-fundo} &&i\frac{\partial}{\partial
T} \Psi_{(0)}(a,T)=\nonumber\\&&\frac{1}{4} \left\{
a^{(3w-1)/2}\frac{\partial}{\partial a} \left[
a^{(3w-1)/2}\frac{\partial}{\partial a}\right] \right\}
\Psi_{(0)}(a,T),
\end{eqnarray}
{for the zeroth order order wave function $\Psi_{(0)}(a,T)$. As we will see in the next section, wave function solutions of this zeroth order equation yield, in the dBB quantum theory, a Bohmian trajectory $a(T)$ through the guidance equations. In this context, the second order equation for the perturbations described by the wave functional $\Psi_{(2)}[v({\bf x}),T]$ can be written as}
\vspace{-12pt}

\begin{eqnarray}
\label{scrhoedinger-separado-perturb}
&&i\frac{\partial}{\partial
T} \Psi_{(2)}[T,v({\bf x})]=
-\frac{a^{(3w-1)}(T)}{2}\int
d^3x\frac{\delta^2}{\delta v^2({\bf x})}\Psi_{(2)}[T,v({\bf x})]+\nonumber\\
&&\frac{w
a^{(3w+1)}(T)}{2}\int d^3x v^{,i}({\bf x})v_{,i}({\bf x})\Psi_{(2)}[T,v({\bf x})].
\end{eqnarray}
Hence, Equation~\eqref{scrhoedinger-separado-perturb} becomes a time-dependent Schr\"odinger equation for $\Psi_{(2)}$ when we substitute $a\rightarrow a(T)$.

One can further perform the time-dependent unitary transformation
\begin{equation}
\label{unitarias} U=\exp\left\{i\int
d^3x\left[\frac{\dot{a}(T)v({\bf x})}{2a(T)}-\frac{(v({\bf x})\pi({\bf x})+\pi({\bf x}) v({\bf x}))}{2}\ln (a(T))\right]\right\},
\end{equation}
yielding
the functional Schr\"odinger equation for the perturbations
\begin{equation}
i\frac{\partial\Psi_{(2)}[v({\bf x}),\eta]}{\partial \eta}= \int \dd^3 x
\left(-\frac{1}{2} \frac{\delta^2}{\delta v^2({\bf x})} +
\frac{w}{2}v_{,i}({\bf x}) v^{,i}({\bf x}) - \frac{{a''}}{2a}v^2({\bf x}) \right)
\Psi_{(2)}[v({\bf x}),\eta], \label{schroedinger-conforme}
\end{equation}
written in terms of conformal time $d\eta=a^{3w-1}d T$ { (the cosmic proper time $\tau$ satisfies $d\tau = a^{3w} dT$; see Section \ref{sec4} for details; hence, $d\eta = d\tau /a$)}, and the new quantum variable $\bar{v}({\bf x})=av({\bf x})$, the usual gauge invariant
Mukhanov--Sasaki variable defined in~\cite{mukh-book} (we have omitted the
bars),

\begin{equation}
\label{vdefinition}
v({\bf {x}})=\frac{{a}^{\frac{1}{2}(3w-1)}}{\sqrt{6}}\biggl( \delta\varphi({\bf {x}})+\frac{2\sqrt{6}\sqrt{(w+1)P_{T}}}{{P}_{a}\sqrt{w}} {a}^{2-3w} {\psi}({\bf {x}})\biggr),
\end{equation}
expressed in terms of the background variables, and the perturbation fields $\psi({\bf {x}}),\delta\varphi({\bf {x}})$. It is connected to the gauge invariant Bardeen
potential $\Phi({\bf x})$ (see~\cite{mukh-book}) through
\begin{equation}
\label{vinculo-simples}
\Phi^{,i}\,_{,i}({\bf x}) =
-\frac{3\sqrt{(\omega+1)\bar{\rho}}}{2\sqrt{\omega}}a
\biggl(\frac{v({\bf x})}{a}\biggr)' ,
\end{equation}

The prime denotes a derivative with respect to conformal time, and $\bar{\rho}$ is the background energy density.

Equation \eqref{schroedinger-conforme} is the usual functional Schr\"odinger equation for quantum linear perturbations in cosmological models with a single perfect fluid satisfying $p=w\rho$. However, the scale factor appearing in Equation~\eqref{schroedinger-conforme} is not the classical one but the Bohmian solution $a(\eta)$. This interpretation of Equation~\eqref{schroedinger-conforme} is only possible within the dBB theory, in which a Bohmian trajectory $a(\eta)$ can be defined. In other frameworks, where $a$ is a background quantum degree of freedom, the physical understanding of the Wheeler--DeWitt Equation~\eqref{schr-fluid} using the ansatz \eqref{ansatz1} implying Equation~\eqref{schroedinger-conforme} becomes conceptually much more intricate, if possible.

The dynamical equation for the quantum operator $\hat{v}({\bf x})$ in the Heisenberg picture~reads
\begin{equation}
\label{ff}
\hat{v}''({\bf x})-\omega \hat{v}^{,i}_{\ ,i}({\bf x})-\frac{a''}{a}\hat{v}({\bf x})=0.
\end{equation}

{The Fourier modes $v_{\bf k}$,}
\begin{equation}
\label{mode}
v({\bf x})=\int{\frac{d^3x}{(2\pi)^{3/2}}v_{\bf k} \ee^{\ii {\bf k} \cdot {\bf x}}},
\end{equation}
{evolve as}
\begin{equation}
\label{equacoes-mukhanov} v''_k+\biggl(\omega
k^2-\frac{{a''}}{a}\biggr)v_k=0.
\end{equation}

These equations have the same form as the equations for scalar
perturbations obtained in~\cite{mukh-book}. However, the function $a(\eta)$
is no longer a classical solution of the background equations but a
quantum Bohmian trajectory of the quantized background. Hence, different power spectra of quantum cosmological perturbations may emerge. In Section \ref{sec5}, we will present the background and quantum perturbation solutions concerning this case.

\subsection{The Canonical Scalar Field}

In the canonical scalar field, the calculations of $H_{(0)}$ and $H_{(2)}$ now yield (without ever using the classical background equations; see~\cite{sandro-scalar} for details),

\be
H_{(0)} = \frac{1}{\ee^{3\al}}\left[ -\frac{\Pi_\alpha^2}{2} + \frac{\Pi_\varphi^2}{2} + \ee^{6\al} V(\varphi) \right],
\label{h00fi}
\en
and

\begin{equation}
\label{h02fi}
H_{(2)}= \frac{1}{2}\int d^3x \left({\pi_v}^2 + {v}^{,i} {v}_{,i}+ 2\frac{z'}{z} {\pi_v}{v}\right),
\end{equation}
where we absorbed $\ka$ in redefinitions of the scalar field and time in order to deal with dimensionless quantities. We set $a=\ee^{\al}$; $v({\bf x})$ is, again, the usual Mukhanov--Sasaki variable~\cite{mukh-book},
\begin{equation}
\label{vinculo-simples2}
v({\bf x}) = a\biggl(\delta\varphi({\bf x}) + \frac{{\varphi} ' \phi({\bf x})}{\cal{H}}\biggr) ,
\end{equation}
with primes denoting derivatives with respect to conformal time $\eta$, $\mathcal{H} = a'/a = \al'$, and $\pi_v$ is its canonical momentum. The background quantities $\Pi_\alpha$ and $\Pi_\varphi$ are the momenta canonically conjugate to background variables $\alpha$ and $\varphi$, respectively, and $N$ plays the role of a Lagrangian multiplier. Finally, $z$ is a background function defined as $z=a\varphi '/\cal{H}$.

As before, when quantizing the system, one obtains the Wheeler--DeWitt equation
\begin{equation}
\label{split-h0-fi}
(\hat{H}_{(0)}+\hat{H}_{(2)}) \Psi = 0,
\end{equation}

Supposing that the background evolution is not affected by the quantum perturbations through some quantum entanglement, one sets

\be
\label{ansatz}
\Psi[\al,\varphi,v({\bf x})] = \Psi_0(\al,\varphi)\Psi_2[\al,\varphi,v({\bf x})].
\en

Note that, in this case, $\Psi_2$ depends on both $\al,\varphi$, as there is no explicit background variable playing the role of time.
The zeroth order part of Equation~\eqref{split-h0-fi},

\be
\label{minibacks1}
\hat{H}_0 \Psi_0 = 0 ,
\en
yields wave solutions that, in the dBB framework, lead to the Bohmian trajectories $\alpha(t)$ and $\varphi(t)$. They will be presented in Section \ref{sec5}.
Having a Bohmian solution for the background, guided by $\Psi_0$, one can now construct the conditional wave equation to describe the perturbations as
\be
\label{cond-fi}
\chi[v({\bf x}),t] = \Psi_2[\al(t),\varphi(t),v({\bf x})].
\en

Using the guidance equations naturally coming from the zeroth order Hamiltonian~\eqref{h00fi} (in the time gauge $N=e^{3\al}$)

\be
\dot \varphi = \pa_\varphi S , \quad \dot \al = - \pa_\al S ,
\label{s6-s}
\en
one obtains

\begin{equation}
\label{essential}
-\left(\frac{\partial S_0}{\partial \alpha}\right)\left(\frac{\partial \Psi_2}{\partial \alpha}\right) +
\left(\frac{\partial S_0}{\partial \varphi}\right)\left(\frac{\partial \Psi_2}{\partial \varphi}\right)=\dot{\al}\left(\frac{\partial \Psi_2}{\partial \alpha}\right) +
\dot{\varphi}\left(\frac{\partial \Psi_2}{\partial \varphi}\right)
= \frac{\partial \chi}{\partial t}.
\end{equation}

Using Equation~\eqref{essential} in \eqref{split-h0-fi}, implementing a time-dependent canonical transformation, similar to what was done in the perfect fluid case (see~\cite{felipe-scalar}), together with one assumption that I will describe soon, one obtains the
Schr\"odinger equation

\begin{equation}
\label{xo2}
\ii \frac{\partial \chi(v,\eta)}{\partial \eta} = \frac{1}{2}\int d^3x \left[ \hat{\pi}^2 + \hat{v}^{,i}\hat{v}_{,i}+ \frac{z'}{z} \left( \hat{\pi}\hat{v}+ \hat{v}\hat{\pi}\right)\right] \Psi(v,\eta),
\end{equation}

Going to the Fourier modes $v_{\bf k}$ of the Mukhanov--Sasaki variable,
\begin{equation}
\label{mode}
v({\bf x})=\int{\frac{d^3x}{(2\pi)^{3/2}}v_{\bf k} \ee^{\ii {\bf k} \cdot {\bf x}}},
\end{equation}
one obtains the mode equation

\begin{equation}\label{eqv2}
v_k''+ \left(k^2-\frac{z''}{z}\right)v_k \quad =0 \qquad .
\end{equation}

As in the perfect fluid case, Equation (\ref{eqv2}) has the same form as the usual equations for the modes in classical backgrounds, but now, the background time functions present in it are the Bohmian trajectories. This can give rise to different effects in the region where the quantum effects
on the background are important, which can propagate to the classical~region.

Equation \eqref{xo2} was obtained under the hypothesis that there is never quantum entanglement between background and the perturbations. When the background behaves classically, this seems to be correct, as the semi-classical calculations, which rely on this hypothesis, yield the observed spectra of perturbations. When the background is quantum, around the bounce, there is nothing imposing the absence of quantum entanglement during this period. In this case, the assumption relies on simplicity. Note that it would be quite interesting to relax the hypothesis of the absence of quantum entanglement around the bounce and investigate its observational consequences.

Finally, I would like to emphasize that, in the case of the scalar fields, there is no degree of freedom that emerges as a possible clock in the original Wheeler--DeWitt equation; see Equations~(\ref{h00fi}),~(\ref{h02fi}) and (\ref{split-h0-fi}). Nevertheless, I was able
to construct a Schr\"odinger equation for the perturbations using the conditional wave function \eqref{cond-fi}. The assumption
of the existence of a Bohmian background quantum trajectory was essential for achieving this goal; see Equation~(\ref{essential}). {The procedure is analogous to what is performed in semi-classical quantum gravity~\cite{kieferSC}, where a notion of time emerges from a combination of the classical background variables (from an equation similar to (\ref{essential})), yielding a background clock, and a Schr\"odinger functional equation for the quantum non-gravitational degrees of freedom is obtained. Within the dBB approach, this can also be performed for a quantum background.} Hence, this is a concrete example, with physical implications, of what was discussed in Section \ref{sec2}. The original Wheeler--DeWitt equation does not have a Shr\"odinger form; it has, rather, a Klein--Gordon form; hence, no notion of probability naturally emerges.
However, in the dBB approach, this difficulty is not an insurmountable obstacle to further calculations, as the wave functional leads to the Bohmian trajectories for the background through the
guidance relations. These trajectories, which are assumed to be actual trajectories, can then be used to construct the conditional wave function for the perturbations,
yielding a Schr\"odinger equation for them. In this case, there is a typical probability distribution~\cite{duerr}, the Born distribution, which can also be the dynamical attractor of any reasonable probability distribution, at a cross-grained level, which is called the quantum equilibrium distribution (see~\cite{valentini}). Consequently, we get back to the standard quantum theory of cosmological perturbations, described by a wave functional with a probabilistic interpretation, but now, the mode perturbations evolve
in a background, the Bohmian background quantum trajectory, which does not always satisfy the background classical GR equations. For other approaches to quantum perturbations in quantum backgrounds, see~\cite{lqcp1,lqcp2,lqcp3}.

\section{Quantum Bouncing Backgrounds and Their Cosmological Perturbations}\label{sec4}

In this section, I will present some examples of background Bohmian solutions that are free of singularities and the features of their cosmological perturbations. I will focus on two matter contents: perfect fluids and a canonical field with an exponential potential.

Perfect fluids can model, quite well, the hot Universe, especially a radiation fluid with $w\approx 1/3$, which usually dominates this very hot phase, not only because of the massless fields present but also because the massive particles have their rest energy completely negligible at high temperatures. Another possibility is that the Universe becomes so dense that the sound velocity of the fluid becomes close to the speed of light, the so-called stiff matter. A canonical scalar field can represent this state if its dynamics are such that its potential energy becomes negligible with respect to its kinetic energy, yielding $p \approx \rho$. This is the case of the exponential potential, which also has other nice properties, as we will see.

\subsection{The Perfect Fluid}

In Section \ref{sec3}, I obtained the Schr\"odinger equations for the background wave function and the background wave functional:

\begin{eqnarray}
\label{scrhoedinger-separado-fundo2} &&i\frac{\partial}{\partial
T} \Psi_{(0)}(a,T)=\nonumber\\&&\frac{1}{4} \left\{
a^{(3w-1)/2}\frac{\partial}{\partial a} \left[
a^{(3w-1)/2}\frac{\partial}{\partial a}\right] \right\}
\Psi_{(0)}(a,T),
\end{eqnarray}
and

\begin{equation}
i\frac{\partial\Psi_{(2)}[v({\bf x}),\eta]}{\partial \eta}= \int \dd^3 x
\left(-\frac{1}{2} \frac{\delta^2}{\delta v^2({\bf x})} +
\frac{w}{2}v_{,i}({\bf x}) v^{,i}({\bf x}) - \frac{{a''}}{2a}v^2({\bf x}) \right)
\Psi_{(2)}[v({\bf x}),\eta], \label{schroedinger-conforme2}
\end{equation}
yielding the normal mode $v_k$ equation
\begin{equation}
\label{equacoes-mukhanov1} v''_k+\biggl(\omega
k^2-\frac{{a''}}{a}\biggr)v_k=0.
\end{equation}

Let us first solve the zeroth order equation. The guidance equations are
\be
\label{guidancec} {\dot T} = \frac{N}{a^{3w}} , \qquad \dot{a}=-\frac{N}{2a} \frac{\partial S}{\partial a}.
\en

Note that, as the resulting Bohmian trajectories have objective reality, the characterization of singularities is very simple and direct, as in classical cosmology: it appears when $a(T)=0$, when space shrinks to zero. {Note, also, that I am treating $T$ in the same way as $a$, with its own guidance equation, and without fixing $N$ from the beginning. The guidance equation for $T$ implies that $dT = N dt/a^{3w}=d\tau /a^{3w}$, where $\tau$ is proper cosmic time. Usually, the time gauge is fixed before quantization, by choosing $N=a^{3w}$, yielding $T=t$ and $d\tau=a^{3w} d T$; hence, both methods are equivalent. As we will see in the sequel, combining both guidance equations of Equation~\eqref{guidancec} yields the same guidance equation for $a$ in terms of $T$, independently of $N$, as it would be obtained if we had fixed $N$ a priori.}

The dynamics can be simplified using the transformation
\be
\label{transf}
\chi=\frac{2}{3} (1-\omega)^{-1} a^{3(1-\omega)/2},
\en
to obtain
\begin{equation}
\ii\frac{\partial\Psi(\chi,T)}{\partial T}= \frac{1}{4} \frac{\partial^2\Psi(\chi,T)}{\partial \chi^2} \label{es202}.
\end{equation}

This is the time-reversed Schr\"odinger equation for a one-dimensional free particle with mass $2$ constrained to the positive axis. As it has a Schr\"odinger form, it is possible, in this case, to obtain the Born rule for $\Psi$ if one imposes the condition

\begin{equation}
\label{cond27}
\Psi \bigl|_{\chi=0} = c \frac{\partial\Psi}{\partial
\chi}\Biggl|_{\chi=0},
\end{equation}
with $c$ being a real constant. For these wave functions, one can assert that $|\Psi^2| d\chi$ is the probability measure for the scale factor, as the boundary condition imposes that the total probability is preserved in time. Another good property of condition \eqref{cond27} is that the Bohmian trajectories coming from wave functions satisfying it are free of singularities~\cite{falciano-santini} because the probability flux $J_\chi \sim {\textrm{Im}}\left(\Psi^* \frac{\pa \Psi}{\pa \chi}\right)$ associated with these wave functions is null at $\chi=0$, so no trajectories can cross $a=0$.
Note that, in the dBB theory, it is not necessary to have a probabilistic interpretation for $\Psi$; hence, one could work with wave functions that do not satisfy boundary condition \eqref{cond27}. In this case, singularities may be obtained, as in plane wave solutions, where the Bohmian trajectories are always classical, hence containing a singularity.

A wave function that satisfies condition (\ref{cond27}) can be obtained by imposing that, at $T=0$, it is the Gaussian
\begin{equation}
\label{initial}
\Psi^{(\mathrm{init})}(\chi)=\biggl(\frac{8}{T_b\pi}\biggr)^{1/4}
\exp\left(-\frac{\chi^2}{T_b}\right) ,
\end{equation}
where $T_b$ is the constant variance of the Gaussian. The wave solution for all times in terms of $a$ reads~\cite{qc2}:
\begin{eqnarray}
\label{psi1t}
\Psi(a,T)&=&\left[\frac{8 T_b}{\pi\left(T^2+T_b^2\right)}
\right]^{1/4}
\exp\biggl[\frac{-4T_b a^{3(1-\omega)}}{9(T^2+T_b^2)(1-\omega)^2}\biggr]
\nonumber\\
&\times&\exp\left\{-\ii\left[\frac{4Ta^{3(1-\omega)}}{9(T^2+T_b^2)(1-\omega)^2}
+\frac{1}{2}\arctan\biggl(\frac{T_b}{T}\biggr)-\frac{\pi}{4}\right]\right\}.
\end{eqnarray}

Taking the two equations in \eqref{guidancec}, one can write a guidance equation describing the dynamics of the scale factor in terms of $T$,
\be
\frac{da}{dT} = - \frac{a^{3w-1}}{2} \frac{\pa S}{\pa a}
\en
or
\be
\frac{d \chi}{dT} = - \frac{1}{2} \frac{\pa S}{\pa \chi}.
\en

Substituting the phase $S$ of (\ref{psi1t}) in these guidance equations, one obtains the Bohmian trajectories
\begin{equation}
\label{at} a(T) = a_b
\left[1+\left(\frac{T}{T_b}\right)^2\right]^\frac{1}{3(1-\omega)} .
\end{equation}

This is a bounce solution without singularities for any initial value $a_b \neq 0$. It tends to the classical solution when $T/T_b\rightarrow\pm\infty$. Hence, the constant $T_b$ provides the time scale of the bounce and the quantum effects. The solution (\ref{at}) can be obtained for other initial wave functions~\cite{falciano-santini}.

The case $w=1/3$ describes a radiation fluid. Adjusting the free parameters conveniently, the solution \eqref{at} can reach the classical Friedmann evolution at energy scales larger than the nucleosynthesis energy scale, when the standard cosmological model begins to be tested by observations. Hence, it is a sensible cosmological model describing the radiation-dominated era, which is free of singularities.

Nevertheless, a complete cosmological model must also contain a presureless component, describing dark matter and baryons (dark energy will be treated later). This extension was accomplished in~\cite{falciano-santini}, yielding
\begin{equation}
\label{ascalefactor}
a(\eta_s)=a_0\left(\dfrac{\Omega_{m0}}{4}\,\eta_s^{2} + \sqrt{\dfrac{1}{x_b^{2}}+\Omega_{r0}\,\eta_s^{2}}\right),
\end{equation}
where $x_b=a_0/a_b$, $a_0$ is the scale factor today, $a_b$ is the scale at the bounce, and $\Omega_{m0}$ and $\Omega_{r0}$ are the usual
dimensionless densities ($\Omega = \rho/\rho_c$) of dust and radiation, respectively, where $\rho_c$ is the critical density.
I have also introduced the dimensionless conformal time, $\eta_s=(a_0/R_{H_0})\eta$, where $R_{H_0}=1/H_0$ is the Hubble radius
today. The scale factor in Equation~\eqref{ascalefactor} describes a universe
dominated by dust in the far past, which contracts up to radiation
domination. Near the singularity, quantum effects become
relevant, and a quantum bounce takes place, eliminating the singularity. The universe is then launched to an expanding phase, reaching the usual standard classical radiation and dust phases. As we will see, the presence of dust is important not only for completeness but also because it is necessary to yield a scale invariant spectrum of scalar perturbations.

The curvature scale at the bounce reads
\begin{equation}\label{lb}
\left. L_{b} \equiv \dfrac{1}{\sqrt{R}}\right\vert_{\eta_s=0} = \left.\sqrt{\dfrac{a^{3}}{6a''}}\right\vert_{\eta_{s}=0},
\end{equation}
where $R$ is the Ricci scalar. It cannot be very close to the Planck length, because near these very small scales, a complete theory of quantum gravity~\cite{qg1,qg2,qg3,qg4} must be evoked. The Wheeler--DeWitt quantization we are using should be understood as a good effective theory for quantum gravity only at higher length scales. Hence, using the values $H_{0}=70
\,\text{km\,\,s}^{-1}\,\text{Mpc}^{-1}$
and $\Omega_{r0}\approx8\times 10^{-5}$, the imposition that $L_b$ should be a few orders of magnitude larger than the Planck length implies that $x_{b} < 10^{31}$. Additionally, as mentioned above, the bounce should
occur before the beginning of the nucleosynthesis era,
implying that $x_{b}\gg 10^{11}$. Collecting these two limits yields
\begin{equation}\label{xblimit}
10^{11}\ll x_{b} < 10^{31}.
\end{equation}

Let us now calculate the mode solutions characterizing the cosmological perturbations of Equation~\eqref{equacoes-mukhanov1}

\begin{equation}
\label{equacoes-mukhanov2}
v''_k+\biggl(\omega
k^2-\frac{{a''}}{a}\biggr)v_k=0,
\end{equation}
in the quantum bounce background given in Equation~\eqref{at}.
Far from the bounce, when $|T|\gg |T_b|$, Equation~(\ref{equacoes-mukhanov2}) reads,

\begin{equation}
\label{Modes} v_k'' +\left[ \omega k^2
+\frac{2(3\omega-1)}{(1+3\omega)^2\eta^2}\right]v_k = 0.
\end{equation}
The solution is
\begin{equation}
\label{Bessel} v_k = \sqrt{|\eta|} \left[ c_1(k) H^{(1)}_\nu
(\bar{k}|\eta|)+ c_2(k)
H^{(2)}_\nu(\bar{k}|\eta|)\right],
\end{equation}
with
$$ \nu = \frac{3(1-\omega)}{2(3\omega+1)}, $$ where $H^{(1,2)}$ are Hankel
functions, $\bar{k}\equiv \sqrt{\omega}k$, and we are considering the far past of the contracting phase, $\eta \ll -1$. In order to obtain spectrum predictions from the above result, one needs to select one solution from Equation~\eqref{Bessel} by fixing $c_1(k)$ and $c_2(k)$.

\begin{figure}[H]

\includegraphics[width=.9\columnwidth]{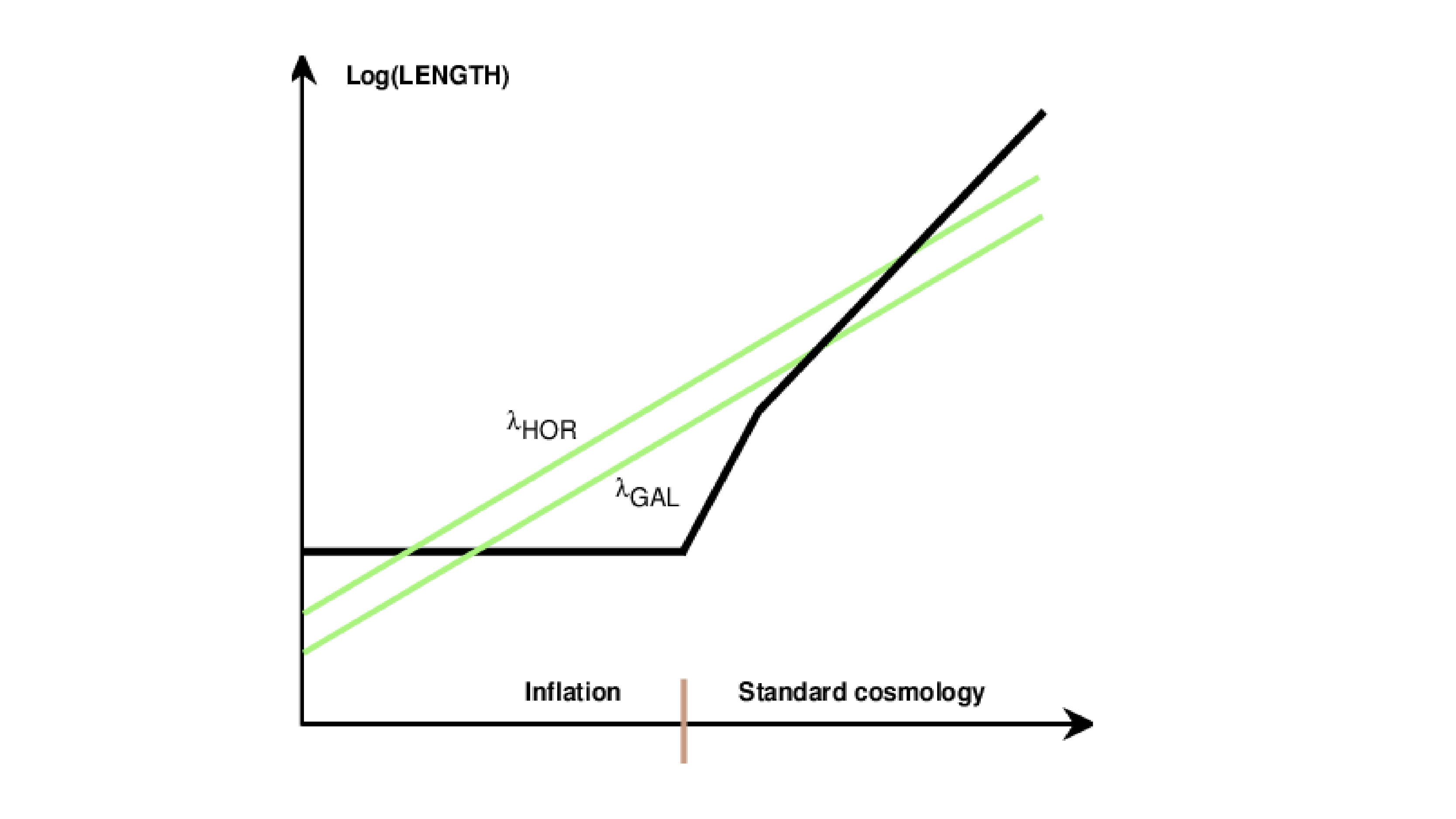}
\caption{Comparison between evolution %Please replace with a sharper image.
 of Hubble radius and cosmological scales in inflation. The green straight lines are the perturbation scales; the black solid line is the Hubble radius. The horizontal axis depicts $\log (a)$.} \label{inf}
\end{figure}

In the case of inflationary models, all the wavelengths of cosmological interest were much smaller than the Hubble radius at least by $60 e-$folds before the end of inflation; see Figure~\ref{inf}. Long before that, any perturbation around the homogeneous background was deep inside the Hubble volume, and it would rapidly fade away, justifying that only quantum vacuum fluctuations could survive. Hence, an adiabatic vacuum state (close to the Bunch--Davies de Sitter vacuum state) is chosen as the initial quantum state of quantum cosmological perturbations. During cosmic evolution, these perturbation scales become bigger than the Hubble radius before and during re-heating, becoming smaller than the Hubble radius again in the expanding decelerating phase. The power spectrum is calculated, with results that remarkably agree with observations~\cite{CMB}. Note, however, that the transition from the quantum description to the classical evolution giving rise to the classical structures in the Universe is very subtle and controversial, needing a clear explanation. This will be done in the next section in the context of the dBB quantum theory.

\begin{figure}[H]

\includegraphics[width=.7\columnwidth]{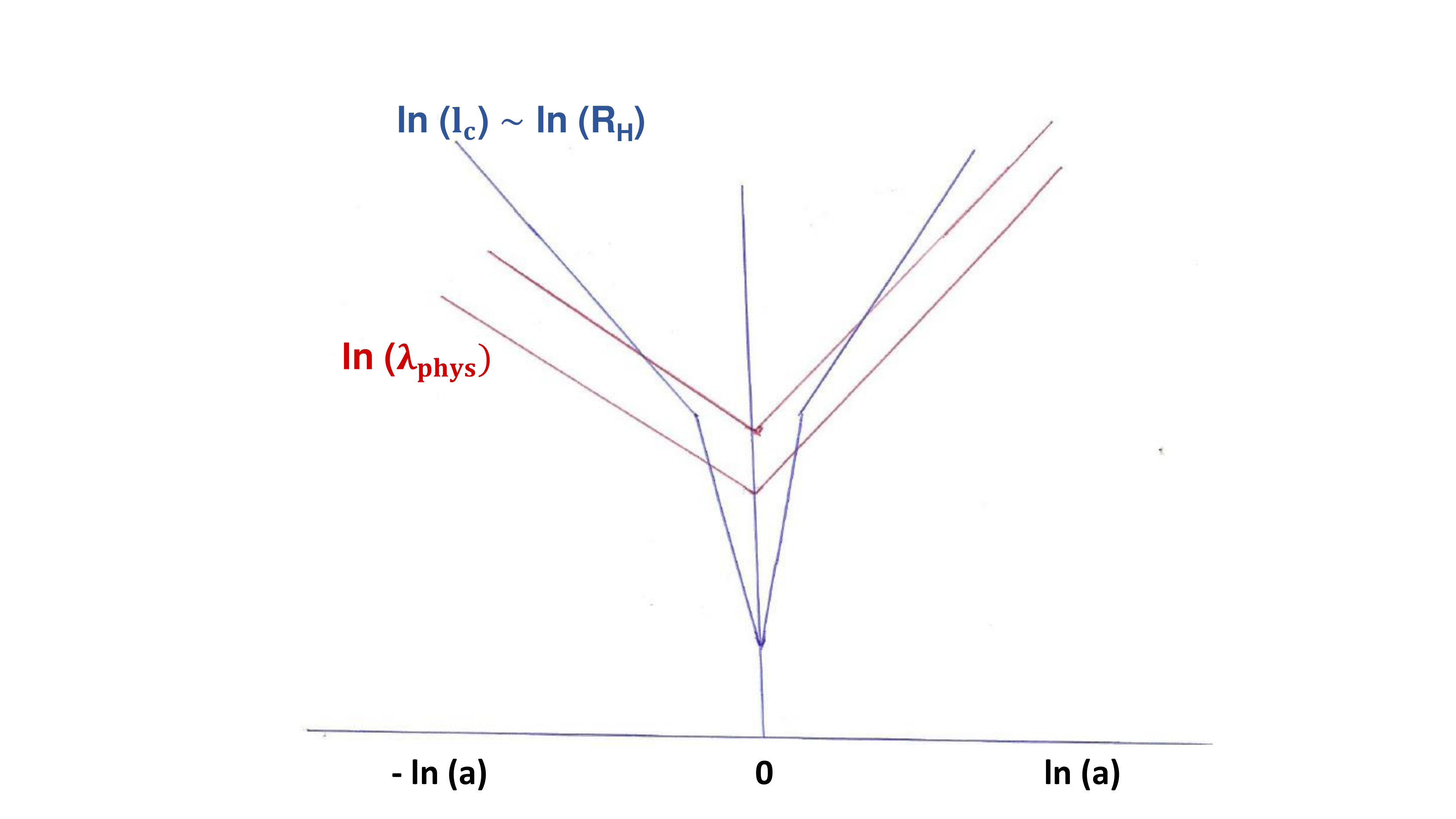}%MDPI: please change the hyphen into minus in the figure
\caption{{ Qualitative comparison between evolution of the curvature scale, in blue, and cosmological scales, in red, in bouncing models. During classical evolution, the curvature scale coincides with the Hubble radius. The scale factor at the bounce is normalized to one; hence, the origin corresponds to the bounce, where the scale factor attains its minimal value. The negative and positive horizontal axis correspond to the contracting and expanding phases, respectively. In the plot, the transitions from dust to radiation domination, and the bounce itself, are qualitatively depicted by sharp transitions. In reality they are smooth, but it does not alter the physical conclusions presented in the text.}} \label{bou}
\end{figure}

The cosmic evolution of the quantum Bohmian bouncing solutions obtained above is completely different from the inflationary solution. As we have seen, they contain a long-standing decelerating contracting phase, implying that they are naturally free of the particle horizon and flatness issues, which are smoothly connected through a quantum bounce to the usual radiation-dominated expanding phase of the standard cosmological model, as can be seen from solution \eqref{ascalefactor}. However, the qualitative evolution of the perturbation scales is very similar. In a decelerating contracting phase, cosmological scales evolve as $\lambda_{\rm phys}\equiv \lambda a \propto \tau^{2/[3(1+w)]}$, while the sound Hubble radius evolves as $R^{S}_{H} = w^{1/2}/H \propto \tau$, where $\tau$ is proper cosmic time, and $H=a'/a^2$ is the Hubble function. Hence, in the far past of the contracting phase, $|\tau|H_0 w^{-1/2}\gg1$, $H_0$ being the Hubble function today; all the scales of cosmological interest were much smaller than the sound Hubble radius as long as $2/[3(1+w)]<1$, or $w>-1/3$, which is exactly the condition for a deceleration. {Put another way, expressing these quantities in terms of the scale factor, one finds that the sound Hubble radius (or the Hubble radius itself) evolves as $R_H \propto a^{3(1+w)/2}$, while the physical cosmological scales are $\lambda_{\rm phys} = \lambda a$. For $w>-1/3$, the Hubble radius grows faster with $a$ than the cosmological scales, implying that, in the far past of the contracting phase of such bouncing models, the cosmological scales were deep inside the Hubble radius. The cases of cosmological interest are dust, radiation and the cosmological constant, in which the exponents appearing in $R_H \propto a^{3(1+w)/2}$ are $3/2, 2, 0$, respectively. In the course of cosmic evolution during contraction, the perturbations scales will eventually become larger than the sound Hubble radius before the bounce. Near the bounce, however, the Hubble radius is no longer a good geometrical scale to which to compare the cosmological scales, as long as the Hubble radius diverges at the bounce, by definition. Indeed, from Equation~\eqref{equacoes-mukhanov2}, one can see that the really physically important geometrical scale to which the cosmological scales must be compared is proportional to the curvature scale $l_c \equiv R^{-1/2}$ ($wk^2\approx a''/a \Rightarrow \lambda_{\rm phys}^2 = a^2/k^2 \approx w a^3/a''=w/R\equiv w l_c^2$), where $R$ is the Ricci scalar of the background. The curvature scale generally coincides with the Hubble radius during classical contraction and expansion, but they behave very differently during the bounce. In fact, the curvature scale has a smooth behavior, without ever diverging. \mbox{Figure \ref{bou}} shows a qualitative comparison between the cosmological scales $\lambda_{\rm phys}\equiv \lambda a$ and the curvature scale $l_c$ in a bouncing model dominated by dust and radiation, plotting $\ln(l_c)$ and $\ln(\lambda_{\rm phys})$ against $\ln(a)$. I normalized $a$ such that the scale factor at the bounce is $1$ ($a_b=1$). The negative (positive) horizontal axis corresponds to the contracting (expanding) phase, respectively. During classical evolution, the curvature scale coincides with the Hubble radius, but it behaves differently during the quantum bounce. Note that the cosmological scales are much smaller than the curvature scale in the far past of the contracting phase; they become larger than the curvature scale during contraction at different times, and they become smaller than the curvature scale again only in the expanding phase. During the period when they are larger than the curvature scale, the perturbations become amplified, yielding the structures in the Universe, as we will see.}
Hence, one can say that decelerating contraction and the bounce play the role of the accelerating phase and re-heating in inflationary models; compare Figure~\ref{inf} with Figure~\ref{bou}. Furthermore, as the sound Hubble sphere in the far past of the contracting phase contains an immensely large space volume and a tiny matter energy density, and as the Universe evolves very slowly because the Hubble time scale $1/H$ is very large, the effective physical universe that affects such perturbation scales is very close to the flat Minkowski space-time. Consequently, any small classical perturbation around this homogeneous, almost-flat background would rapidly dissipate away, and, as in inflation, only quantum vacuum fluctuations would survive (see~\cite{novo}), justifying, again, the choice of an adiabatic vacuum state as the initial quantum state of quantum cosmological perturbations, which is now close to the Minkowski vacuum quantum state. Hence, the qualitative justification for imposing vacuum initial conditions for the cosmological perturbations in inflation and bouncing models is similar, although the physical ambiences justifying them are completely different. Note, however, that the quantum perturbations in bouncing models must be dynamically evolved through a different background, especially through the bounce, which generally involves new physics, with possible different observational consequences, as we will see.

The modes characterizing a(n) (almost) Minkowski vacuum state are given by

\begin{equation}
v_{k}^{(\mathrm{ini})} =
\frac{\exp {i \bar{k} \eta}}{\sqrt{\bar{k}}}.
\label{v}
\end{equation}

The asymptotic expansion of the Hankel functions for $k|\eta|\gg 1$
that fits solution \eqref{Bessel} with~\eqref{v} implies that
$$ c_1=0 \quad \hbox{and} \quad c_2= l_{P} \sqrt{\frac{3\pi}{2}}
\exp^{-i\frac{\pi}{2} \left(\nu+\frac{1}{2}\right)}.
$$

In order to propagate the solution through the bounce up to the expanding phase, one expands the solutions of Equation~\eqref{Modes} in powers of $k^2$ according to
the formal solution~\cite{mukh-book}
\begin{eqnarray}
\frac{v_k}{a} & = & A_1(k)\biggl[1 - \omega k^2 \int^{\eta} \frac{d\bar
\eta}{a^2\left(\bar \eta\right)} \int^{\bar{\eta}}
a^2\left(\bar{\bar{\eta}}\right)d\bar{\bar{\eta}}\biggr]\nonumber
\\ &+& A_2(k) \biggl[\int^\eta\frac{d\bar{\eta}}{a^2} - \omega k^2
\int^\eta \frac{d\bar{\eta}}{a^2} \int^{\bar{\eta}} a^2
d\bar{\bar{\eta}} \int^{\bar{\bar{\eta}}}
\frac{d\bar{\bar{\bar{\eta}}}}{a^2} \biggr] +\;...\;,\cr & & \label{solform}
\end{eqnarray}
where I have omitted the terms of order $\mathcal{O}(k^{j\geq 4})$. The quantity $v/a$ is the curvature perturbation $\zeta$~\cite{mukh-book}. This solution is adequate when $wk|\eta|\ll 1$. The solution \eqref{Bessel} is also valid in this regime, as long as the bouncing solution \eqref{at} is still in the classical regime. This is true for all scales of cosmological interest because they cross the sound Hubble radius, which happens when $wk|\eta|\approx 1$, when the Universe is still very large, and quantum effects are completely negligible.
Hence, in this region, one can match the solution \eqref{solform} with solution \eqref{Bessel}, and obtain
the coefficients $A_1(k)$ and $A_2(k)$ from the coefficients $c_1(k)$ and $c_2(k)$, which were fixed by the vacuum initial condition \eqref{v}. They read

\begin{eqnarray}
A_1 &\propto&
\biggl(\frac{\bar{k}}{k_0}\biggr)^\frac{3\left(1-\omega\right)}
{2\left(3\omega+1\right)},\label{A1}\\ A_2 &\propto& \biggl(\frac{\bar{k}}{k_0}
\biggr)^\frac{3\left(\omega-1\right)}{2\left(3\omega+1\right)},\label{A2}
\end{eqnarray}
where $k^{-1}_0=T_0a_0^{3\omega-1}=L_b$, and $L_b$ is the curvature scale at the bounce. Propagating the solution \eqref{solform} up to the expanding phase, and relating it to the Bardeen potential $\Phi({\bf x})$ through the known formula

\begin{equation}
\label{vinculo-simples2}
\Phi^{,i}\,_{,i}({\bf x}) =
-\frac{3 l_{P}^2\sqrt{(\omega+1)\bar{\rho}}}{2\sqrt{\omega}}a
\biggl(\frac{v({\bf x})}{a}\biggr)' .
\end{equation}
one obtains, in the expanding phase for $T\gg T_b$,

\begin{equation}
\Phi_k \propto
k^\frac{3\left(\omega-1\right)}{2\left(3\omega+1\right)}
\biggl[\mathrm{const.}+\frac{1}{\eta^{(5+3\omega)/(1+3\omega)}}\biggr].
\end{equation}

The constant mode contains a mixing of the coefficients $A_1$ with $A_2$ in the expanding phase, but the $A_2$ coefficient is multiplied by a large constant, dominating over $A_1$; see~\cite{large-sandro}.
Hence, calculating the power spectrum of the Bardeen potential,
\begin{equation}
\label{PS} \mathcal{P}_\Phi \equiv \frac{2 k^3}{\pi^2}
\left| \Phi_k \right|^2,
\end{equation}
which is connected to the anisotropies of the CMBR and fixed by observations, one obtains
\begin{equation}
\mathcal{P}_\Phi \propto k^{n_{_\mathrm{S}}-1},
\label{powspec}
\end{equation}
where
\begin{equation}
\label{indexS} n_{_\mathrm{S}} = 1+\frac{12\omega}{1+3\omega}.
\end{equation}

In the case of gravitational waves, the
equation for the modes $\mu_k = a h_k$, where $h_k$ is the mode related to the amplitude of the wave, can be obtained very easily, because gravitational waves are gauge invariant; see~\cite{PPNGW2} for details. It is given by
\begin{equation}
\label{mu} \mu_k''+\left( k^2 -\frac{a''}{a} \right)\mu_k =0.
\end{equation}

The power spectrum is
\begin{equation}
\label{PT} \mathcal{P}_h \equiv
\frac{2 k^3}{\pi^2}\left| \frac{\mu_k}{a} \right|^2 \propto k^{n_{_\mathrm{T}}},
\end{equation}
and it reads
\begin{equation}
\label{indexT} n_{_\mathrm{T}} = \frac{12\omega}{1+3\omega}.
\end{equation}

One can see from Equation~\eqref{indexS} that, for $\omega\approx 0$ (dust), one obtains a nearly scale-invariant spectrum for both tensor and scalar perturbations~\cite{PPNscalar}. This is a general result for bouncing models~\cite{Allen2004,bounce-classical8,mbounce2,mbounce3}: if the contracting phase of a smooth bouncing model is dominated at large scales by a matter field satisfying $w=p/\rho\approx 0$, then the power spectrum of long wavelength scalar perturbations in the expanding phase is nearly scale invariant. {However, there is a problem if the matter field is a fluid in which $w=c_s^2$ because, in this case, $w$ must be positive and one cannot obtain a red-tilted spectrum, as observed. In the case of a canonical scalar field, this is not the case because $w$ is independent of $c_s^2$, and it can be made negative, as we will see in the next subsection. Nevertheless, one can circumvent this problem even in the fluid case.} Note that it is not
necessary that the dust fluid dominates at all times. As we have seen above, the
$k$-dependence of $A_1$ and $A_2$ is obtained far from the
bounce, when the modes cross the sound Hubble radius, $\bar{k}\eta\approx 1$, and they do not change in a possible transition from matter
to radiation domination in the contracting phase or across a smooth
bounce. The effect of the bounce is to mix the two
coefficients, and the constant mode acquires, in the expanding phase, the scale-invariant piece. Hence, the bounce itself may be dominated by another fluid,
such as radiation. In fact, the more complete bounce solution \eqref{ascalefactor} also yields an almost scale-invariant spectrum of adiabatic cosmological perturbations. Its amplitude reads~\cite{large-sandro,2-fluids}

\begin{equation}
\label{ampl-ad}
A_S \approx 10^{-2} \frac{l_p^2}{R_{H_0}^2} \frac{x_b^2}{\Omega_{r0}c_s^5},
\end{equation}
where $c_s$ is the value of the sound velocity characterizing the adiabatic perturbation when the perturbation scale crosses the sound Hubble radius. Note that the amplitude becomes bigger for small values of the sound velocity. Indeed, the perturbation modes grow faster after they cross the sound Hubble radius, which shrinks if $c_s$ becomes small. Hence, they cross this scale earlier for smaller $c_s$ and have more time to grow.
As $A_S \approx 2.09 \times 10^{-9}$ (see~\cite{CMB}), and using Equation~\eqref{xblimit}, one obtains

\begin{equation}\label{cslimit}
10^{-16} \leq c_s < 10^{-10}
\end{equation}
for the sound velocity. Note,
however, that there are two fluids; hence, the sound velocity for the adiabatic perturbations reads:

\begin{equation}
\label{sound-ad}
c_s^2 = \frac{w(\rho_m + p_m) + (\rho_r + p_r)/3}{\rho_T + p_T},
\end{equation}
where $w$ is the equation of the state parameter of the dust fluid, and the indices $m,r,T$ designate the dust, radiation, and total energy densities and pressures, respectively. Hence, $c_s^2 \approx |w|\ll 1$ only when dust dominates. For small scales that cross the sound Hubble radius very late, near radiation domination, the power spectrum amplitude is highly suppressed because $c_s$ is no longer in the range \eqref{cslimit}, as it tends to increase up to $1/\sqrt{3}$ when radiation begins to be important. Hence, the spectrum must be slightly red-tilted due to the presence of radiation, and the parameters may be fitted with CMBR observations.

\begin{figure}[H]
\includegraphics[width=0.7\textwidth]{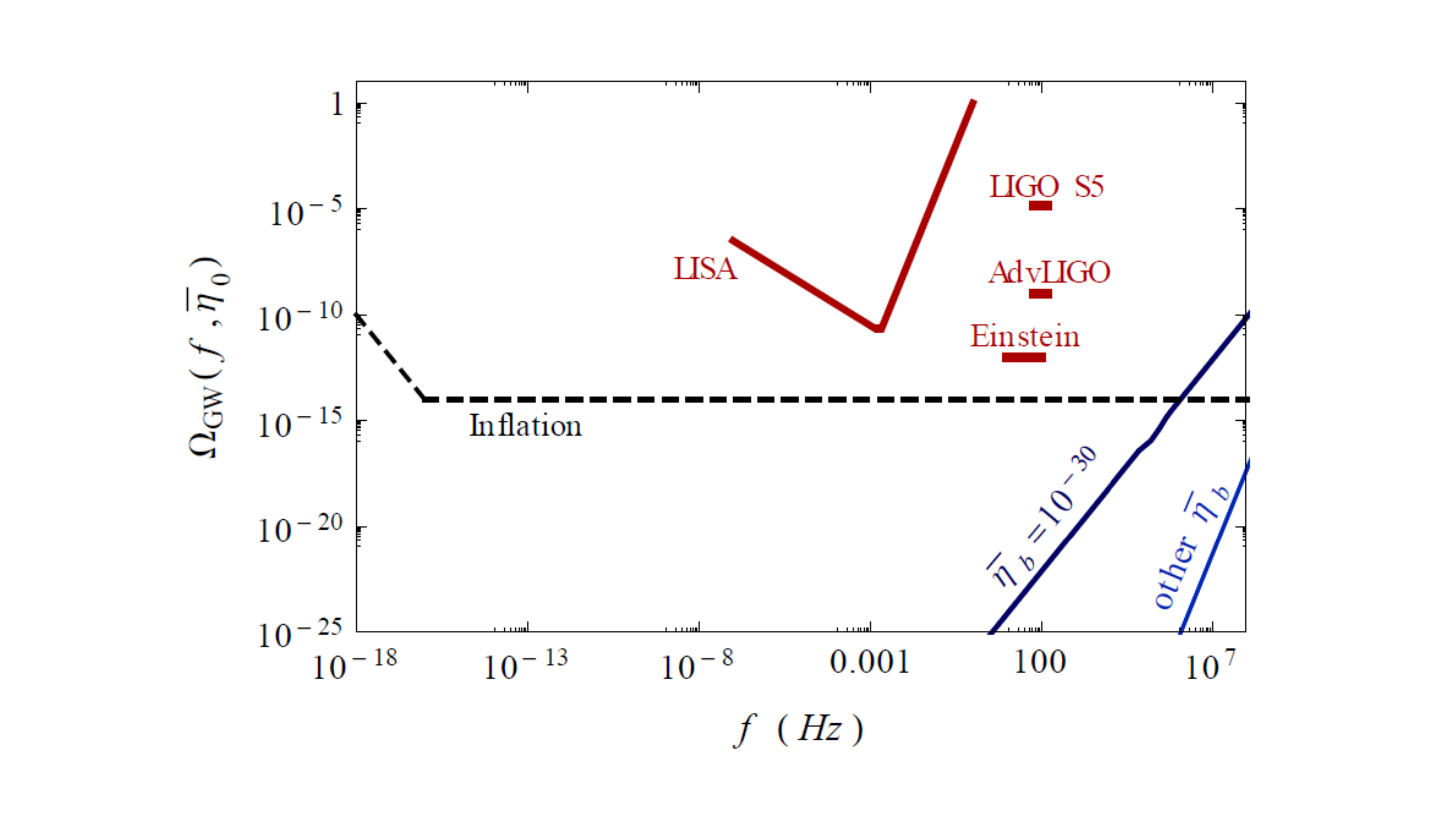}
\caption{The figure shows a comparison of our results, labeled by ${\bar{\eta}}_b$ (the smaller this
parameter, the bigger the energy scale of the bounce, and the value $10^{-30}$ is only two orders of magnitude
away from the Planck scale) with experimental sensitivities of LIGO's 5th run, Advanced LIGO, and the forthcoming LISA and Einstein Telescope, and a prediction of the upper limits on the spectrum of primordial gravitational waves generated in inflationary models.}
\label{GW1}
\end{figure}

Note that, as tensor perturbations have $c_s=1$, and as they evolve similarly to scalar perturbations, their amplitudes must be very small in comparison with scalar perturbations, being unobservable at large scales or very small frequencies. However, they might be observable at larger frequencies. Indeed, we have calculated the strain spectrum of the stochastic background of relic gravitons in such models~\cite{denis}, and we have shown that the resulting amplitude is too small to be detected by any gravitational wave detector, unless in the frequency range 10--100 Hz, as can be seen from Figure~\ref{GW1}. However, it is a hard technical challenge to detect stochastic gravitational waves in such range of frequencies, if possible.

\begin{figure}[H]

\includegraphics[width=.6\columnwidth]{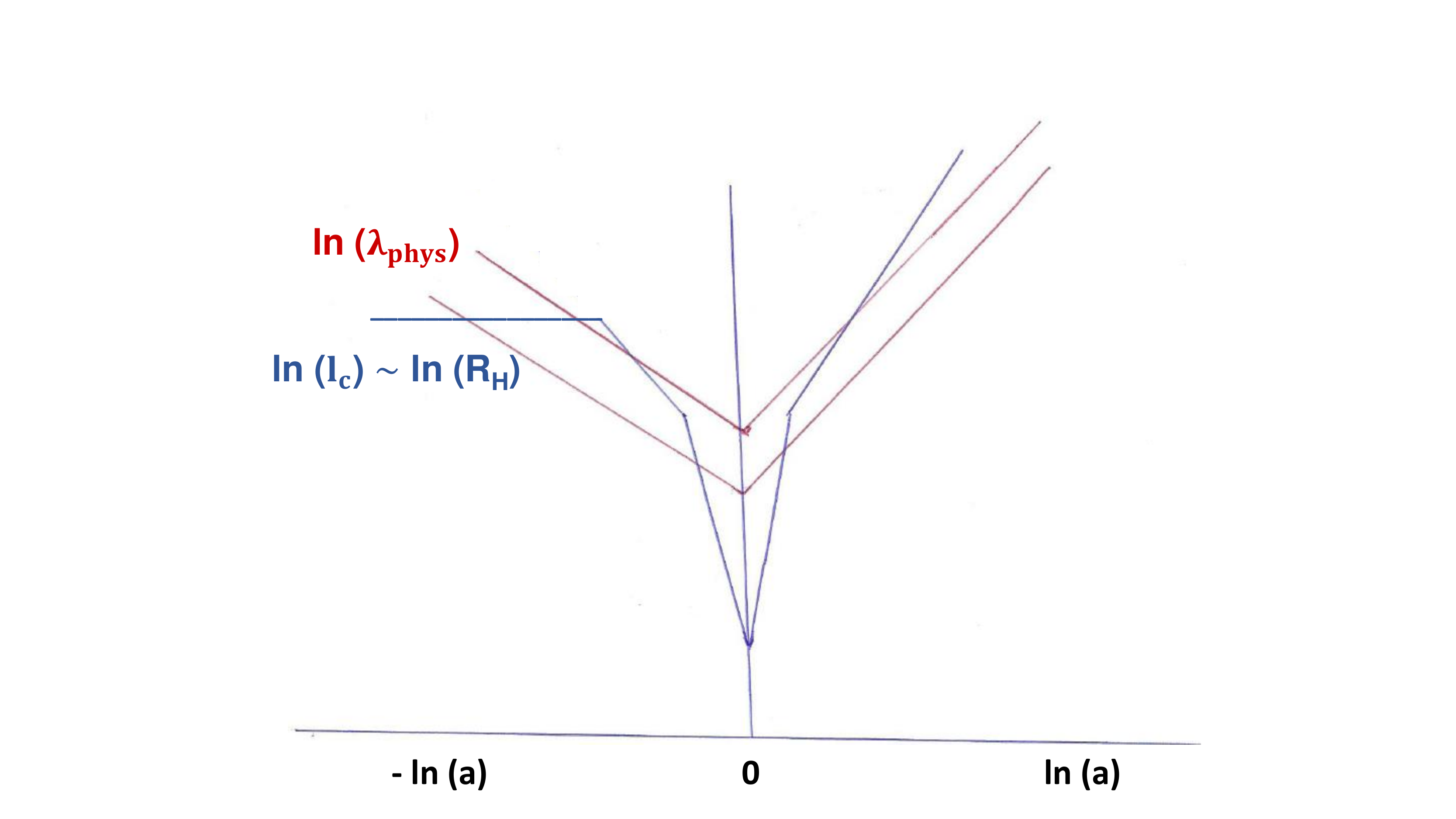}
\caption{{ Qualitative comparison between evolution of the curvature scale, in blue, and cosmological scales, in red, in bouncing models with a cosmological constant. During classical evolution, the curvature scale coincides with the Hubble radius. The scale factor at the bounce is normalized to one; hence, the origin corresponds to the bounce, where the scale factor attains its minimal value. The negative and positive horizontal axis correspond to the contracting and expanding phases, respectively. In the plot, the transitions from dust to radiation domination and dust to cosmological constant domination, and the bounce itself, are qualitatively depicted by sharp transitions. In reality, they are smooth, but it does not alter the physical conclusions presented in the text.}}
\label{Blam}
\end{figure}

\begin{figure}[H]

\includegraphics[width=.8\columnwidth]{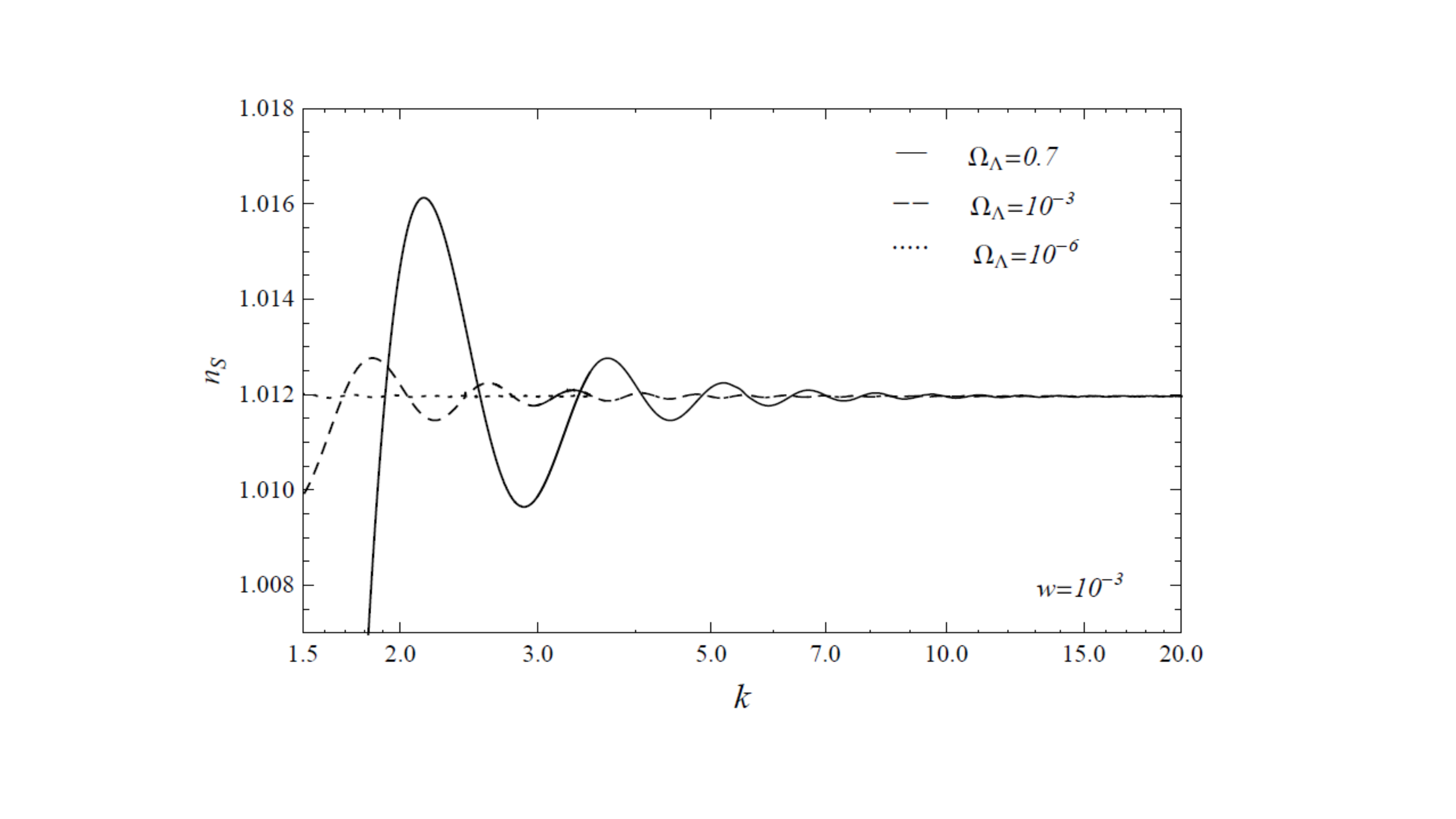}
\caption{Numerical results for $n_S(k)$ in the presence of a cosmological constant. The solid line shows the result obtained using $\Omega_{\Lambda}=0.7$; the dashed line, that for $\Omega_{\Lambda}=10^{-3}$; and the dotted line, that for $\Omega_{\Lambda}=10^{-6}$.
The oscillations become smaller for
smaller $\Omega_\Lambda$, indicating that they arise because of the presence of the cosmological constant.}
\label{fig3}
\end{figure}

Up to now, I have not considered dark energy (DE), which seems to be accelerating the present Universe~\cite{lss,crs}. In the case of inflationary models, DE is irrelevant, because the initial conditions for the perturbations and their subsequent evolution are set at very small scales, where DE does not play any role. However, in bouncing models, vacuum initial conditions for quantum cosmological
perturbations are set in the far past of the contracting phase of these models, when the Universe was very large and almost flat, and DE energy may be relevant at such large scales, as it is in the expanding phase of our Universe. In fact, in the case of the standard $\Lambda$CDM model, where DE is a cosmological constant, the {the curvature scale, or} sound Hubble radius, stops evolving linearly in cosmic time and tends to be constant at large scales. Hence, going back in time, as the large perturbation scales grow following a time power-law, $\tau^{1/[3(1+w)]}$, they become larger than the Hubble radius again, and an adiabatic Minkowski vacuum prescription for their initial conditions become problematic; see Figure~\ref{Blam}. One possible solution to this problem is to try to define a Minkowski adiabatic vacuum in the period of time when the cosmological constant is not relevant,
but the Universe is still very large, with a Hubble radius larger than the scales of cosmological interest. However, these cosmological scales are not much smaller than the length scale
associated with the value of the cosmological constant given in the $\Lambda$CDM standard cosmological model; hence,
the spectrum of cosmological perturbations at these scales can be influenced by its presence.
In fact, we have shown in~\cite{Maier2011}, analytically and numerically, that, in a bouncing model containing a dust fluid ($w\approx 0$) and a cosmological constant, an
almost scale-invariant spectrum of long-wavelength perturbations is also obtained, but it is now affected by the presence of the cosmological constant. It
induces small oscillations and a small running towards a red-tilted spectrum in these scales; see Figure~\ref{fig3}.
Hence, small oscillations in the spectrum of temperature fluctuations may arise in the cosmic background radiation at large scales, superimposed to the usual acoustic~oscillations.

\vspace{-6PT}

In the next subsection, I will present the scalar field case. Concerning primordial gravitational waves, as the sound velocity associated with canonical scalar field scalar perturbations is $1$, the scalar and tensor perturbations evolve approximately in the same way in classical bouncing models, rendering  %Please check intended meaning is retained
the ratio of the tensor to scalar perturbations of order $1$, $r=T/S \approx 1$~\cite{CaiGW1,CaiGW2}, which is ruled out by observations. I will show that, in the case of a quantum bounce, the quantum effects near the bounce increase the scalar perturbations with respect to tensor perturbations, yielding $r < 0.1$, as observed. Furthermore, using an exponential potential, the problem with dark energy mentioned above is circumvented, as we will see.

\subsection{Canonical Scalar Field}

Consider a canonical scalar field $p=X-V(\varphi)$ in which
\begin{equation}
\label{def_pot}
V(\varphi) = V_0 \e^{-\lambda {\bar \kappa} \varphi},
\end{equation}
where $V_0$ and $\lambda$ are constants. ${\bar \kappa}^2 = 6 \kappa^2 = 8\pi G$, so that $\lambda$ is dimensionless.

Exponential potentials have been widely studied in cosmology, as they can model primordial inflation, the present acceleration of the Universe, and matter bounces. Their scalar field dynamics in expanding Friedmann backgrounds contain an attractor where the ratio between the pressure and the energy density is constant: $w=p/\rho$, where $w=(\lambda^2-3)/3$. Hence, by adjusting $V_0$ and $\lambda$, they can be used to describe the above cosmological scenarios.

Restricting ourselves to matter bounces, $w\approx 0$, one must set $\lambda \approx \sqrt{3}$. Note that, as $w$ is not related to the sound speed squared of scalar perturbations, as in the fluid case, it is not restricted to being positive: it can have a small negative value in order to give $n_s = 1 + 12w/(1+3w) \approx 0.97$.

Let us first present the classical dynamics of canonical scalar fields with exponential potential. Using cosmic proper time, $N=1$, one can define the variables
\begin{equation}\label{mud1}
x = \frac{{\bar \kappa} }{\sqrt{6}H}\dot{\varphi}, \qquad
y = \frac{{\bar \kappa} \sqrt{V_M}}{\sqrt{3}H},
\end{equation}
where
\be
H=\frac{\dot a}{a} = {\dot \al}
\en
is the Hubble parameter. This choice dramatically simplifies the Friedmann equations as follows:
\begin{equation}\label{sist1c}
\frac{\dd x}{\dd \alpha} = -3 \left(x-\frac{\lambda}{\sqrt{6}}\right) \left(1-x\right)\left(1+x\right)
\end{equation}

\begin{equation}
\label{xyFried}
x^2 + y^2 = 1.
\end{equation}

The ratio $w = p/\rho$ is given by
\begin{equation}
w = 2x^2-1. \label{x_y_con}
\end{equation}

The critical points of this system are listed in Table~\ref{tab_crit}; see~\cite{heard-wands}. The critical points at $x=\pm 1$, yielding $p=\rho$ (the potential is negligible with respect to the kinetic term), and the scalar field behave as stiff matter. They correspond to the space-time singularity $a=0$. The critical points $x=1/\sqrt{2}$ imply that $w=0$ (see Equation~\eqref{x_y_con}) or $p=0$, and the scalar field behaves as dust matter. They
are attractors (repellers) in the expanding (contracting) phase, corresponding to very large, slowly expanding (contracting) universes, and the space-time is asymptotically flat in time. Additionally, from Equation~\eqref{x_y_con}, one can see that, at $x=0$, the scalar field behaves like dark energy; $w=-1$, $p=-\rho$.
\begin{specialtable}[H]
\tablesize{\small}
\caption{Critical points of the planar system defined by \eqref{sist1c} and \eqref{xyFried}.}
\label{tab_crit}
\setlength{\cellWidtha}{\columnwidth/3-2\tabcolsep+0.0in}
\setlength{\cellWidthb}{\columnwidth/3-2\tabcolsep+0.0in}
\setlength{\cellWidthc}{\columnwidth/3-2\tabcolsep+0.0in}
\scalebox{1}[1]{\begin{tabularx}{\columnwidth}{>{\PreserveBackslash\centering}m{\cellWidtha}>{\PreserveBackslash\centering}m{\cellWidthb}>{\PreserveBackslash\centering}m{\cellWidthc}}
\toprule

\emph{\textbf{x}} & \emph{\textbf{y}} & \emph{\textbf{z}} \\
\midrule
$-1$ & $0$ & $1 $\\
\midrule
$1$ & $0$ & $ 1$ \\
\midrule
$\frac{\lambda }{\sqrt{6}}$ &$ -\sqrt{1-\frac{\lambda ^2}{6}}$ & $\frac{1}{3} \left(\lambda ^2-3\right)$ \\
\midrule
$\frac{\lambda }{\sqrt{6}}$ & $\sqrt{1-\frac{\lambda ^2}{6}}$ & $\frac{1}{3} \left(\lambda ^2-3\right)$ \\
\bottomrule
\end{tabularx}}

\end{specialtable}

The expanding solutions evolve from a Big Bang singularity, when the scalar field behaves as stiff matter, up to an asymptotic future where the scalar field behaves as dust. In one of the possibilities, the scalar field passes through a dark energy phase.

The contracting solutions evolve from an asymptotic past dust-dominated contraction, ending in a Big Crunch singularity, when the scalar field behaves as stiff matter. Again, in one of the possibilities, the scalar field passes through a dark energy phase.

\begin{figure}[H]
\includegraphics[width=0.6\textwidth]{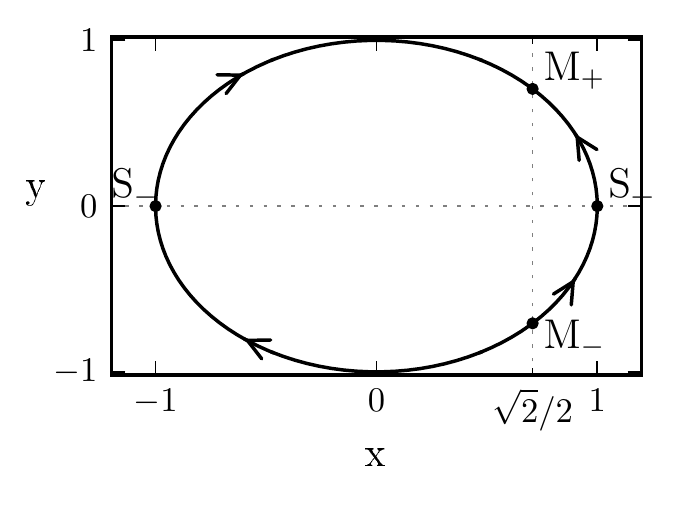}
\caption{Phase space for the planar system defined by \eqref{sist1c} and \eqref{xyFried}. The critical points are indicated by $M_\pm$ for a
dust-type effective equation of state, and $S_\pm$ for a
stiff-matter equation of state. Note that the region $y < 0$ shows the contracting solutions, while the
$y > 0$ region presents the expanding solutions. Lower and upper quadrants are not physically
connected, because there is a
singularity in~between.} \label{phase-classical}
\end{figure}

These classical possibilities are shown in Figure~\ref{phase-classical}. The Friedmann Equation \eqref{xyFried} restricts the trajectories to a circle. The upper and down semicircles are disconnected, as the $S_{\pm}$ points are singularities, and they show the expanding and contracting solutions, respectively. The points $M_{\pm}$ are the dust attractor and repeller points, respectively.

Let us now quantize this background model and perturbations around it, using the results of Section \ref{sec3}.

\begin{figure}[H]
\includegraphics[width=0.6\textwidth]{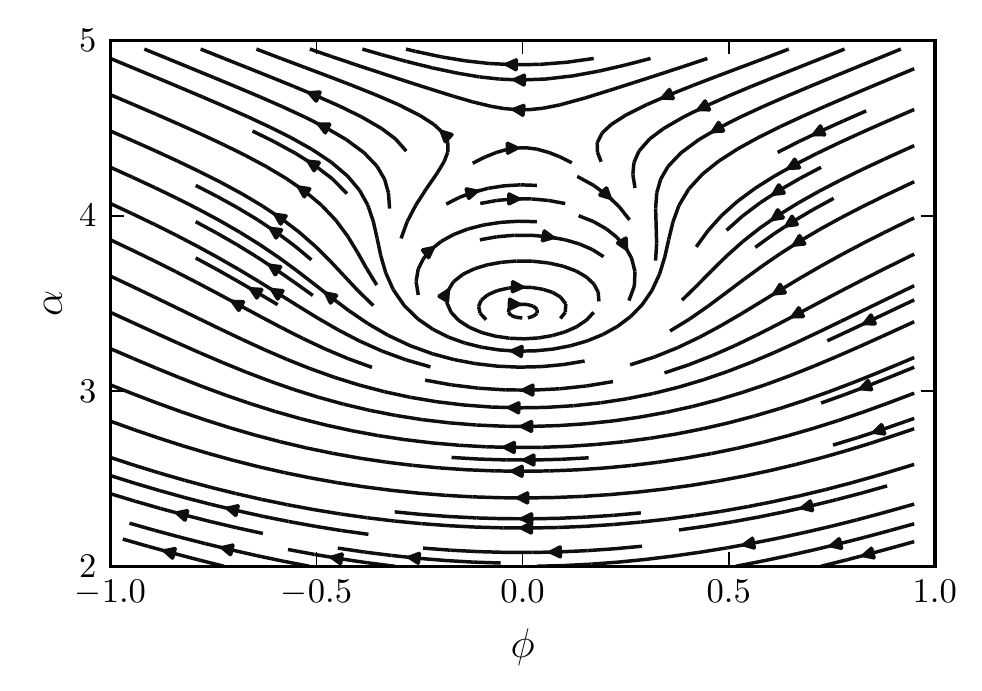}
\caption{Phase space for the quantum bounce~\cite{bacalhau17}. The bounces in the figure connect regions around $S_+$ in the contracting phase with
regions around $S_-$ in the expanding phase.} \label{quantum-bounce}
\end{figure}

One first has to solve the background Wheeler--DeWitt Equation \eqref{minibacks1}, where ${\hat{H}}_{(0)}$ is the operator version of the classical background Hamiltonian

\be
H_{(0)} = \frac{1}{\ee^{3\al}}\left[ -\frac{\Pi_\alpha^2}{2} + \frac{\Pi_\varphi^2}{2} + \ee^{6\al} V(\varphi) \right],
\en
where $V(\varphi)$ is the exponential potential. Exact solutions were found in~\cite{colin17} with their respective Bohmian trajectories, which are non-singular bouncing solutions. These solutions can also be obtained in a more simplified way by noting that, near the singularity, the scalar field behaves as stiff matter, the potential can be neglected, and solutions to this case were found in~\cite{qc7}, whose trajectories around the bounce are shown in \mbox{Figure~\ref{quantum-bounce}}. Note that, for large scale factors, $\alpha \gg 1$, the classical stiff matter behavior is recovered, $x \approx \pm 1$, and from there on, the Bohmian trajectories become classical. They can then be appropriately matched with the classical solutions presented in Figure~\ref{phase-classical}. This was done in~\cite{bacalhau17}, yielding the same qualitative picture.
In fact, both results, together with some general arguments based on the Bohmian configuration space, where no trajectories can cross (see~\cite{colin17}), imply that the only Bohmian possible bounce solutions are those that connect the region around $S_{\pm}$ with the region around $S_{\mp}$. Figure~\ref{quantum-bounce} shows a concrete example of a quantum bounce transiting from $S_+$ to $S_-$. Hence, the possible Bohmian bouncing scenarios are:
\begin{itemize}
\item[(A)]
A long classical dust contraction, which traverses a dark energy phase and realizes a stiff matter quantum bounce, directly expanding afterwards to an asymptotically
dust matter expanding phase, without passing through a dark energy phase.
\item[(B)]
A long classical dust contraction, without traversing a dark energy phase, which realizes a stiff matter quantum bounce and expands to a dark energy phase, ending in an asymptotically
dust expanding phase.
\end{itemize}

Case B is the physically interesting solution. First, it contains a dark energy phase in the expanding era, allowing the description of the present observed acceleration of the Universe. Second, there is no dark energy phase in the contracting era. The model has a long standing dust contraction, where space-time is almost flat in its asymptotic past, allowing the prescription of an adiabatic Minkowski vacuum as the initial state for quantum cosmological perturbations and avoiding the problem concerning the quantum vacuum prescription for the initial quantum state of cosmological perturbations when dark energy is present. Hence, the dBB quantum theory yields an example where vacuum initial conditions for quantum cosmological perturbations can be easily imposed in bouncing models with dark energy. {Other physical effects can lead to bouncing models with a dark energy phase~\cite{leh,caiU,od}, with similar properties. The relevant aspect of the present model is that a single canonical scalar field, with a quite simple potential, was capable of modeling not only a dark energy phase at large scales in the expanding phase of the bouncing model but also a pressureless field that dominates the asymptotic past of the contracting phase of the same model.}

Having solved the background Wheeler--DeWitt Equation~\eqref{minibacks1} and found the relevant background Bohmian trajectories, let us now calculate the amplitudes of scalar perturbations and primordial gravitational waves in this background.

As we have seen in Section \ref{sec3}, quantum primordial gravitational waves are described by the variable $\mu$, whose modes satisfy similar equations to the Mukhanov--Sasaki variable mode $v_k$, with the scale factor $a$ playing the role of $z$:

\begin{equation}
\label{mode-v2}
v_{\bf k}'' + \left(k^2-\frac{z''}{z}\right) z_{\bf k}=0 ,
\end{equation}

\begin{equation}
\label{mode-f2}
\mu_{\bf k}'' + \left(k^2-\frac{a''}{a}\right) \mu_{\bf k}=0 .
\end{equation}
Adiabatic vacuum initial conditions are set using the mode Equation~\eqref{v}, which applies for both $v_k$ and $\mu_k$. The calculations in the Bohmian background were performed in~\cite{bacalhau17}.

In order to qualitatively understand the final results, let us discuss what happens near the quantum bounce. Approaching the bounce in the contracting phase, the perturbation reaches the super-Hubble behavior, where $z''/z \gg k^2$ and $a''/a \gg k^2$. As shown in Section~\ref{sec3}, in this regime, the solutions for the scalar and tensor perturbations at leading order in a $k^2$ expansion read
\begin{align}\label{solHa}
{\zeta}_k \equiv \frac{v_k}{z} &\approx A^{(1)}_k + A^{(2)}_k \frac{1}{R_H}\int \frac{\dd{}\tau}{x^2a^3}, \\
{h}_k \equiv \frac{\mu_k}{a} &\approx B^{(1)}_k + B^{(2)}_k \frac{1}{R_H}\int \frac{\dd{}\tau}{a^3},
\end{align}
where $x$ was defined in Equation~\eqref{mud1}.

\begin{figure}[H]

\includegraphics[width=.7\columnwidth]{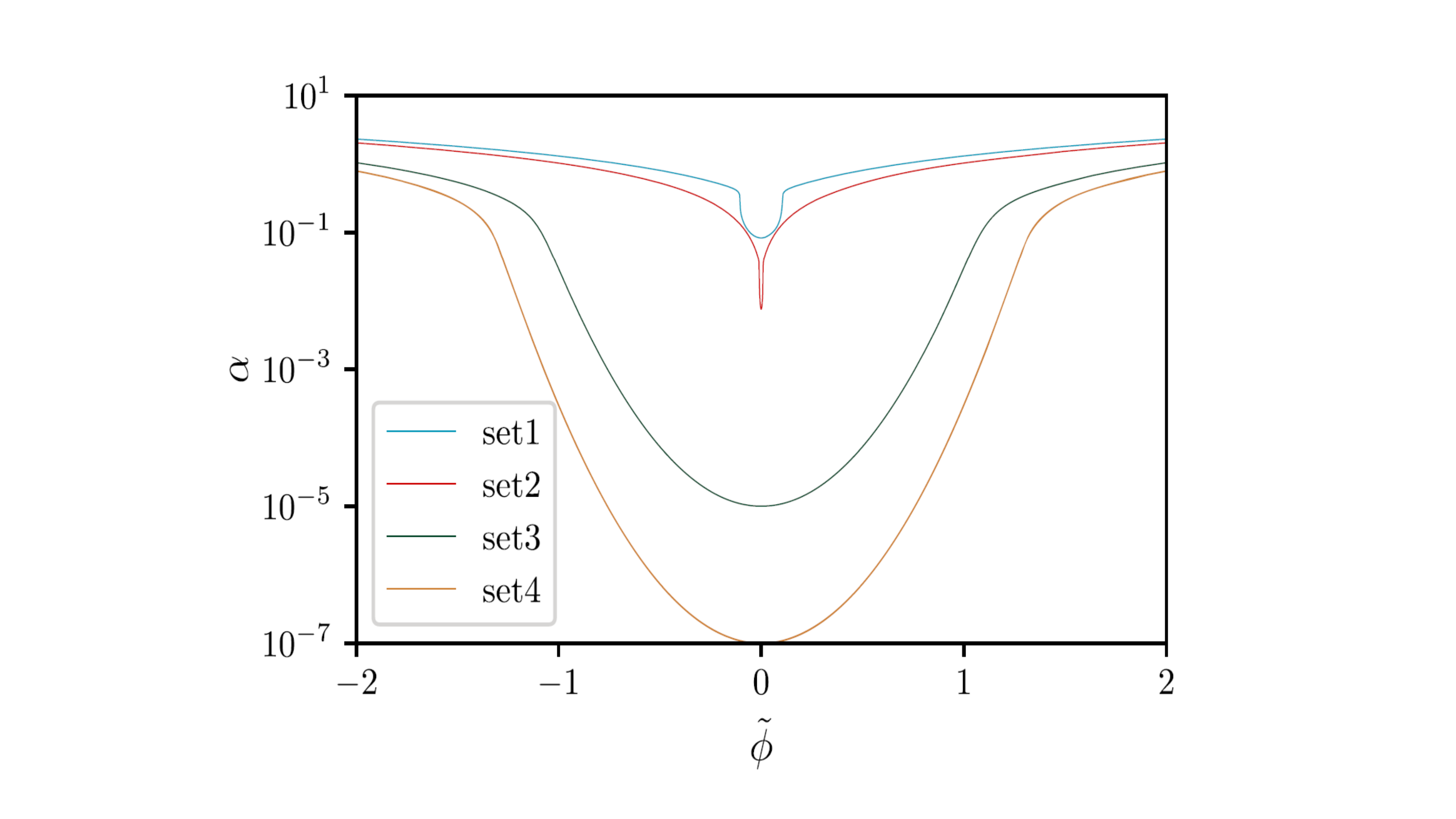}
\caption{Possible Bohmian trajectories associated with the canonical scalar field with exponential potential. The trajectories yielding relevant amplification of scalar perturbations are set 1 and set 2. The bounces are not deep, but they are steep, with very small $x$.} \label{quantum-bounce2}
\end{figure}

\begin{figure}[H]

\includegraphics[width=0.6\textwidth]{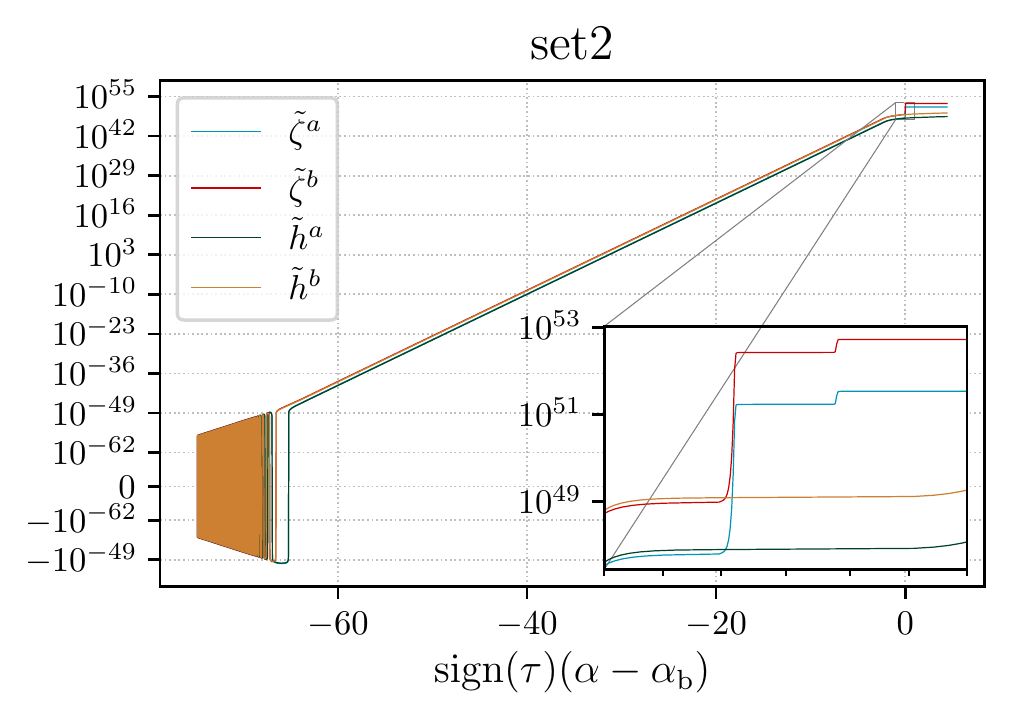}

\caption{Evolution of scalar and tensor perturbations in the background of case B. Scalar
and tensor perturbations grow almost at the same rate during classical contraction, but
at the quantum bounce, the scalar perturbations are enormously enhanced over the tensor perturbations
due to the quantum effects (shown in the detail of the figure). After the bounce, the perturbations
get frozen. The final amplitudes of both perturbations are compatible with observations. The indices
$a$ and $b$ refer to the real and imaginary parts of the perturbation amplitudes.}
\label{zeta_h}
\end{figure}

In the classical contracting phase of case B, one has $0< x< 1/\sqrt{2}$; hence, the evolution of $\zeta_k$ and $h_k$ is very close, since they are different by the presence of $x$ in Equation~\eqref{solHa}, implying that $r = T/S \approx 1$. This is the origin of the problem with classical bouncing models with canonical scalar fields. In a quantum bounce, however, the classical Friedmann equations %Please check intended meaning is retained
are no longer satisfied, the evolution is no longer restricted to the circle in Figure~\ref{phase-classical}, and $x$ can assume any value. Indeed, in Figure~\ref{quantum-bounce2}, one can see that there are Bohmian trajectories where $x=d\varphi/d\alpha$ is very small. Hence, in this period, the scalar perturbation amplitudes can increase relatively to the tensor perturbation amplitudes. Indeed,
this was calculated numerically, and the results are shown in Figure~\ref{zeta_h}. One can see a sharp increase in the scalar perturbation amplitude
around $|\alpha-\alpha_b| \approx 10^{-1}$, where $\al_b$ is the value of the scale factor at the bounce.

\vspace{-12pt}

This is a remarkable result. It shows that features of quantum Bohmian trajectories can lead to observational consequences and explain involved cosmological issues, such as the unwanted large ratios of tensor to scalar
perturbation amplitudes that plague classical bouncing models with canonical scalar fields. Hence, it is a concrete example of how a quantum cosmological effect can be amplified to yield sound observable consequences.

The free parameters of the theory can be adjusted to yield the right amplitudes and
spectral indices of scalar and tensor perturbations. {From Planck observations, one obtains $n_s = 0.9652\pm 0.0042$, implying that $\lambda^2 = 2.9914 \pm 0.0010$. The amplitude of scalar perturbations, $A_s \approx 2.1 \times 10^{-9}$, can be obtained if the curvature scale at the bounce is around $10^3 l_p$. However, the bounce must be steep in order to obtain sufficient amplification of scalar perturbations over tensor perturbations, as shown in Figure~\ref{zeta_h}; see Figure~\ref{quantum-bounce2} and~\cite{bacalhau17} for details.} Hence, applying the dBB quantum theory to quantum cosmology made it possible to obtain a simple and sensible bouncing model with dark energy behavior in the expanding phase, and
correct and well-defined perturbation amplitudes of quantum mechanical origin.

\section{The Quantum-to-Classical Transition of Quantum Cosmological Perturbations}\label{sec5}

As we have seen in Section \ref{sec3}, in both bouncing and inflationary models, the seeds of structure in the Universe are the quantum
fluctuations of an adiabatic vacuum, which is defined when the wavelengths of cosmological interest are deep inside the sound Hubble radius, either in the slow contracting phase of a very large and rarefied universe in the far past of bouncing models or in the quasi-de Sitter expansion of inflationary cosmology. During the evolution of
the Universe, these quantum vacuum fluctuations must become classical fluctuations, as the structures present in the real universe (galaxies, cluster of galaxies, etc.) are classical.

In the context of the Copenhagen interpretation, it is rather difficult, if not impossible,
to explain this transition. The adiabatic vacuum is a homogeneous and isotropic quantum state. For instance, the mean value of the curvature perturbation $\zeta(x)$ squared in the adiabatic vacuum state $|0\rangle$ is homogeneous:

\begin{equation}
\langle0|\zeta^2(x)|0\rangle = \langle0|T^{\dagger}\zeta(x)T T^{\dagger}\zeta(x)T |0\rangle=\langle0|\zeta^2(x+\delta)|0\rangle,
\end{equation}
where $T$ is the $\delta$ translation unitary operator, and the vacuum state satisfies $T|0\rangle = |0\rangle$.

From another point of view, the temperature anisotropies that are measured in the CMBR~\cite{CMB} originated from the Sachs--Wolff effect are obtained from the Bardeen potential $\Phi$, but how should we understand $\Phi$ in this calculation: as a mean value of the quantum operator corresponding to $\Phi$ (which is zero in the state $|0\rangle$) or as a particular realization of it? Which one?

The usual attempts to address these issues argue that, in inflation, the vacuum state is squeezed, yielding a positive Wigner distribution in phase space, which looks like a classical stochastic distribution of realizations of the Universe, with different inhomogeneous configurations. Decoherence avoids interference among the different realizations.
These arguments were severely criticized by many authors~\cite{lyth,mukh-book,weinberg-book,sudarsky}. The quantum state, although squeezed, is still homogeneous, so what breaks its homogeneity? What is the environment of the perturbations in the decoherence picture? The most fundamental question is as follows: in the Copenhagen interpretation, different potentialities are not realities, so how does one of the potentialities become our real Universe? How does a single outcome emerge from the many possible realizations, without a collapse of the wave function, or what defines the role of a measurement in the early Universe? This main issue is ultimately connected with the measurement problem in quantum mechanics, which, as commented on in the Introduction, becomes acute when the physical system is the Universe under the Copenhagen view. It cannot be solved by the arguments above without a collapse postulate, which does not make sense in the physical situation we are facing: we cannot collapse the perturbation wave function because we could not exist without stars!

The dBB quantum theory provides a simple and elegant solution to this very important problem. First of all, remember that, besides the quantum state, there is an actual quantum field describing the cosmological perturbations, which, depending on its initial configuration, will distinguish one of the possible realizations of the Universe with respect to the others, breaking the symmetry of the quantum state. As explained in Section \ref{sec2}, there is no collapse, but one realization is selected by the evolution of the actual perturbation field. Secondly, as be shown in the sequel, this quantum evolution becomes classical while the Universe evolves, either in inflation or in bouncing models.

As we have seen in Section \ref{sec3}, the Schr\"odinger equation for the perturbations reads

\begin{equation}
\label{xo2}
\ii \frac{\partial \Psi(v,\eta)}{\partial \eta} = \frac{1}{2}\int d^3x \left[ \hat{\pi}^2 + \hat{v}^{,i}\hat{v}_{,i}+ \frac{z'}{z} \left( \hat{\pi}\hat{v}+ \hat{v}\hat{\pi}\right)\right] \Psi(v,\eta),
\end{equation}
where $z=a\varphi '/\cal{H}$ is a background function coming from either an inflationary model or a quantum Bohmian trajectory (which may be nonsingular with a bounce), as we will see in the next section.

Going to the Fourier modes $v_{\bf k}$ of the Mukhanov--Sasaki variable,
\begin{equation}
\label{mode}
v({\bf x})=\int{\frac{d^3x}{(2\pi)^{3/2}}v_{\bf k} \ee^{\ii {\bf k} \cdot {\bf x}}},
\end{equation}
and because of linearity, one can set the product wave function
\be
\label{product}
\Psi= \Pi_{{\bf k} \in \setR^{3+}} \Psi_{\bf k}(v_{\bf k},v^*_{\bf k},\eta),
\en
where each factor $\Psi_{\bf k}$ satisfies the Schr\"odinger equation
\begin{equation}
\label{sch}
\ii\frac{\partial\Psi_{\bf k}}{\partial\eta}=
\left[ -\frac{\partial^2}{\partial v_{\bf k}^*\partial v_{\bf k}}+
k^2 v_{\bf k}^* v_{\bf k}
- \ii\frac{z'}{z}\left(\frac{\partial}{\partial v_{\bf k}^*}v_{\bf k}^*+
v_{\bf k}\frac{\partial}{\partial v_{\bf k}}\right)\right]\Psi_{\bf k}.
\end{equation}

The guidance equations are
\begin{equation}
\label{guidance}
v'_{\bf k}= \frac{\partial S_{\bf k}}{\partial v^*_{\bf k}}+\frac{z'}{z}v_{\bf k} .
\end{equation}

The wave function $\Psi_{\bf k}$ associated with the adiabatic vacuum wave functional given by Equation~\eqref{product} reads (see~\cite{polarski} for details)
% start a new page without indent 4.6cm

\end{paracol}

\begin{equation}
\label{psi2}
\Psi_{\bf k} = \frac{1}
{\sqrt{\sqrt{2\pi}|f_k(\eta)|}} \exp{\left\{-\frac{1}{2|f_k(\eta)|^2}|v_{\bf k}|^2 + i \left[\left(\frac{|f_k(\eta)|'}{|f_k(\eta)|}-
\frac{z'}{z}\right)|v_{\bf k}|^2-
\int^\eta \frac{d {\tilde \eta}}{2|f_k({\tilde \eta})|^2}\right]\right\}} ,
\end{equation}

\begin{paracol}{2}
%\linenumbers
\switchcolumn
where $f_k$ is a solution of the classical mode equation
\begin{equation}
\label{mode-f}
f_{\bf k}'' + \left(k^2-\frac{z''}{z}\right) f_{\bf k}=0,
\end{equation}
with initial conditions $f_k(\eta_i) = \exp{-i k\eta_i}/\sqrt{2k}$, where $|\eta_i|$ is an early time in the contracting phase where $k^2 \gg z''/z$ \footnote{In this section, I will name the Mukhanov--Sasaki mode $v_k$ of Section \ref{sec4} $f_k$, reserving the name $v_k$ for the Bohmian mode that we are now discussing.}. This state is homogeneous and isotropic. Note that, around $\eta_i$, the wave function $\Psi_k$ reduces to the usual harmonic oscillator ground state wave function for the mode $k$, with ground state energy $E_k=k$

The guidance equations can be integrated to give
\be
\label{soly}
v_{\bf k}(\eta) = v_{\bf k}(\eta_i)\frac{|f_k(\eta)|}{|f_k(\eta_i)|},
\en
independently of the particular form of $f_k(\eta)$.

When $k^2 \gg z''/z$, we have seen that, in either inflation or the bounce scenario in the contracting phase, the solution of Equation~\eqref{mode-f} can usually be approximated to

\begin{equation}
\label{fka}
f_k(\eta) \sim \ee^{- \ii k \eta}\left(1 + \frac{A_k}{\eta} + {\rm {O}}(\eta^{-2}) + \dots \right).
\end{equation}

In the simple solutions presented in Section \ref{sec3}, Equation~\eqref{fka} comes from asymptotic Hankel function expansions.
By inserting Equation~\eqref{fka} into Equation~\eqref{soly}, one obtains, for the Bohmian modes,
\begin{equation}
v_{\bf k}(\eta) \sim \left( 1+\frac{{\rm Re} A_k}{\eta} + \dots \right).
\end{equation}

Note that $v_{\bf k}$ is approximately constant, as it is the usual case of Bohmian trajectories corresponding to the ground state of a harmonic oscillator. Hence, the Bohmian mode is completely different from the classical mode: the first is almost static, and the second is oscillating. The quantum perturbation field is genuinely quantum.

When the modes get deep inside the potential, $k^2 \ll z''/z$, we learned from Section IV that the classical mode $f_{\bf k}$ is a combination of power law solutions, which soon becomes dominated by a growing mode. Hence, one has
\begin{equation}
\label{yqq}
f_k (\eta) \sim A_k \eta ^{\beta},
\end{equation}
where $\beta<0$ \footnote{In inflation, this result is direct, while for bouncing models, some care must be taken with the interchange between growing and decaying modes after the bounce, but in the end, the result is the same; see~\cite{qtc2}.}. Hence, as $|f_k|$ equals $f_k$, up to a time-independent complex factor and, looking at Equation~\eqref{soly}, the Bohmian modes evolve in the same way as the classical modes in this era. The classical limit has been achieved, long before non-linear structures begin to be formed.

One can also use the quantum potential to investigate the classical limit, constructing it from the wave function Equation~\eqref{psi2} in both eras. It was explicitly shown in~\cite{qtc1,qtc2} that, indeed, when $k^2 \gg z''/z$, the quantum potential dominates the evolution of the perturbations, while for $k^2 \ll z''/z,$ it becomes negligible with respect to the classical potential.

In order to obtain the statistical prediction, one can write the Bohmian field as $v(\eta,{\bf x};v_i)$ such that $v(\eta_i,{\bf x};v_i) = v_i({\bf x})$. If the initial field $v_i$ is distributed according to the quantum equilibrium distribution $|\Psi(v_i,\eta_i)|^2$, we have seen that $v(\eta,{\bf x};v_i)$ will be distributed according to $|\Psi(v,\eta)|^2$. This property is called equivariance. For such an equilibrium ensemble, we can consider the two-point correlation function
\begin{eqnarray}
\label{2-point}
&&\left\langle v(\eta,{\bf x})v(\eta,{\bf x}+{\bf r})\right\rangle_{\rm B} \nonumber \\
&=& \int \mathcal{D} v_i |\Psi(v_i, \eta_i)|^2 v(\eta,{\bf x};v_i) v(\eta,{\bf x} + {\bf r};v_i) \nonumber \\
&=& \int \mathcal{D} v |\Psi(v, \eta)|^2 v({\bf x}) v({\bf x} + {\bf r}).
\end{eqnarray}

The second line expresses the integration over the ensemble of possible initial configurations with distribution $|\Psi(v_i,\eta_i)|^2$, and the step to the third line is a consequence of equivariance.

Using Equations~(\ref{product}) and (\ref{psi2}), one obtains

\begin{equation}
\label{twopoint}
\left\langle {v}({\bf x},\eta) {v}({\bf x} + {\bf r},\eta)\right\rangle_B = \frac{1}{2\pi^2}\int k^2 d{k} \frac{\sin{kr}}{kr} |f_{k}(\eta)|^2 \equiv \frac{1}{2\pi^2}\int d\ln{k} \frac{\sin{kr}}{kr} P(k,\eta),
\end{equation}
where $P(k,\eta) = k^3 |f_{k}(\eta)|^2$ is the power spectrum of $v$.
Note that

\begin{equation}
\label{two-2}
\left\langle {v}({\bf x},\eta) {v}({\bf x} + {\bf r},\eta)\right\rangle_B=\left\langle \hat{v}({\bf x},\eta) \hat{v}({\bf x} + {\bf r},\eta)\right\rangle,
\end{equation}
where the R.H.S is the usual two-point correlation function in the Heisenberg representation, calculated in Section \ref{sec4}.

One can object that Equation~\eqref{2-point} is an average over possible realizations of the Universe and we just see one universe. This can be overcome with the argument that the width of the Gaussian distribution of temperature correlations in the CMBR is small for small angles; {hence, a measurement related to these small correlation angle temperature--temperature anisotropies must be very close to the mean value. Note that, for larger angles (or larger cosmological scales), this is no longer the case, leading to the so-called cosmic variance (a larger imprecision in these larger angle observations).} For details, see~\cite{mukh-book}.

In conclusion, the dBB approach explains, in a very simple and clear way, the quantum-to-classical transition of quantum cosmological perturbations, conceptually and qualitatively, solving an ancient deep problem concerning the evolution of cosmological perturbations of quantum mechanical origin. For other approaches and other quantum effects in the CMBR, see~\cite{cmbq1,cmbq2,cmbq3,cmbq4,cmbq5}.

\section{Discussion and Conclusions}\label{sec6}

As we have seen in this review, de Broglie--Bohm (dBB) quantum theory is very
suitable for quantum cosmology. Many of the issues that plagued the subject for a long time simply disappear:

\begin{itemize}
\item [(1)] The measurement problem is naturally solved, without the necessity of invoking the presence of an external agent outside the quantum physical system, which does not make sense when the physical system is the whole Universe.

\item[(2)] The fact that the usual quantum equations for the wave function of the Universe that emerge from many approaches to quantum gravity do not present a Schr\"odinger form makes it difficult to physically interpret the wave function of the Universe, especially in probabilistic terms~\cite{kuchar,kuchar2}. In the dBB theory, however, the wave function of the Universe $\Psi$ yields the guidance equations, which provide the time evolution of all the quantum particles and fields present in the Universe. Hence, one can assign a nomological interpretation to $\Psi$, as giving the laws of motion for the quantum degrees of freedom, in the same way as Hamitonians and Lagrangians do. There is no need to talk about probabilities at this level; hence, the quantum equations for $\Psi$ may have any form. When dealing with subsystems in the Universe, one can construct the conditional wave function to describe this subset of fields and particles, which may satisfy a Schr\"odinger-like equation under reasonable assumptions, and a natural probabilistic interpretation in terms of the Born rule emerges for this conditional wave function.

\item[(3)] There is the so-called problem of time in quantum cosmology, as it seems that the quantum theory is timeless~\cite{kuchar2}. This issue is intimately connected with the second one. In the dBB quantum approach, the guidance equations yield a parametric evolution for the fields. Note, however, that the space-time structure that emerges from orderly stacking the fields along this parameter may be very contrived, but they can be calculated; see~\cite{pinto-neto1} for details. Additionally, when going to subsystems described by the conditional wave function, where a Schr\"odinger-like equation emerges, a time evolution for the subsystem quantum state emerges.

\item[(4)] As, in the dBB theory, the Bohmian trajectories emerge, the characterization of quantum singularities becomes clear. For instance, in the quantum cosmological models discussed here, the background model is said to be non-singular if the Bohmian trajectory of the scale factor satisfies $a(t)\neq 0$ for all $t$.

\item[(5)] The classical limit is easily obtained, either by the inspection of the quantum potential or by direct comparison between the classical and Bohmian trajectories.
\end{itemize}

The features of the dBB theory, which naturally solve the issues of quantum cosmology presented above, have many important consequences:

\begin{itemize}

\item[(i)] Feature (1) yields a clear understanding of a long standing problem, which is the quantum-to-classical transition of quantum cosmological perturbations in inflation and bouncing models. This was discussed in Section \ref{sec5}.

\item[(ii)] Feature (4) allows a simple identification of non-singular quantum models, as shown in Section \ref{sec4}. All of them present a regular bounce.

\item[(iii)] All the features above yield simple equations for quantum perturbations in quantum backgrounds, which is not an easy task under other approaches~\cite{halli}. These simple equations could be solved, providing sensible bouncing models with inhomogeneous perturbations, in which the presence of a dust fluid (dark matter?) yields an almost scale-invariant spectrum of perturbations, as observed, with the correct amplitudes. Dark energy can also be included, as in the scalar field model of Section \ref{sec4}. In this model, we have seen that a quantum cosmological effect becomes very relevant during the quantum bounce, leading to observable consequences which solve a conflict with observational results that cannot be solved in classical terms, rendering it a viable model to be developed.

\item[(iv)] Feature (5) makes direct the evaluation of the parameter limits under which the standard classical Friedmann solution arises from a quantum Bohmian solution.

\end{itemize}

There are many routes of investigation to be deepened, and many new to be followed. Some of them are:

\begin{itemize}

\item[(a)] The angular power spectrum of the temperature--temperature correlation function, and the $E$ and $B$ polarization modes corresponding to the bouncing models described here, and other possibilities, must be calculated in great detail, and compared with the most recent CMB results~\cite{planck}, in order to differentiate these models among themselves and with inflation. Additionally, one could try to find typical fingerprints of a quantum cosmological effect, which cannot be found by other methods. One promising example is the scalar field model presented in Section \ref{sec4}.

\item[(b)] In the analysis of more elaborate models, some new observables must be calculated. For instance, in the two-fluid model described in Section \ref{sec5}, one needs to calculate the entropy perturbations. In preliminary calculations~\cite{2-fluids}, as the entropy effective sound velocity is given by

\begin{equation}
\label{sound-ad}
c_e^2 = \frac{w(\rho_r + p_r) + (\rho_m + p_m)/3}{\rho_T + p_T},
\end{equation}
the large scale perturbations become super-Hubble in the dust-dominated era, when $c_e^2=1/3$, and the short scale perturbations in the radiation-dominated era, when $c_e^2=w\approx 0$. Hence, in opposition to the adiabatic perturbations, the large scale entropy perturbations are very small compared to adiabatic perturbations, which is in agreement with observations, but they may be relevant at small scales. Hence, even knowing that small scale perturbations are suppressed by Silk damping, some imprint of this effect may be present. Furthermore, they may also slightly affect the spectrum index of large scale adiabatic perturbations. This is work in progress.

\item[(c)] The role of dark energy in bouncing models is very important to understand. In Section \ref{sec5}, I presented a possible solution to the issues raised in Section \ref{sec4}, but there are many other possibilities. In the case that dark energy is a cosmological constant, the problem becomes more contrived, with the possibility of observational consequences. Note that bouncing models offer a unique possibility to learn about dark energy through the primordial power spectrum of cosmological perturbations, which is not the case for inflation.

\item[(d)] The effects of a quantum bounce on non-gaussianities are also a very relevant investigation, with possible observational consequences~\cite{cai2,agullo}.

\item[(e)] The dBB quantum theory, in principle, allows probability distributions that do not obey the Born rule, that are away from quantum equilibrium. It is difficult to find ordinary physical systems in this situation. In cosmology, this may not be the case. For instance, long wavelength perturbations originated from a vacuum quantum state do not relax quickly to quantum equilibrium, yielding a possible departure from
quantum mechanical predictions~\cite{PVV}. Additionally, one could relax the conditions imposed in the conditional wave function explained in Section V, which would lead to corrections to the effective Schr\"odinger equation for the perturbations in the quantum background regime and a departure from quantum
equilibrium, with possible observational consequences.

\end{itemize}

In conclusion, quantum theory can indeed help cosmology in solving the singularity problem, which plagues all GR classical solutions. Furthermore, in a reverse way, cosmology can also help in understanding quantum theory more deeply. For instance, the alternative Many Worlds Interpretation~\cite{eve} was constructed because the Copenhagen interpretation cannot be applied to quantum cosmology. Additionally, we have seen that another alternative, the dBB quantum theory, yields possible testable predictions and new effects that may distinguish it from other quantum approaches. These possible tests appear only in a cosmological context. Hence, not only does quantum theory help cosmology but, also, cosmology can improve quantum theory.

Let me end with a Louis de Broglie quote~\cite{undivided}:

\begin{quote}
“To try to stop all attempts to pass beyond the
present viewpoint of quantum physics could be
very dangerous for the progress of science and
would furthermore be contrary to the lessons
we may learn from the history of science.
This teaches us, in effect, that the actual state
of our knowledge is always provisional
and that there must be, beyond what is actually
known, immense new regions to~discover”.
\end{quote}
\vspace{6pt}

\funding{This research received no external funding.}%: ``This research received no external funding'' or ``This research was funded by NAME OF FUNDER grant number XXX.'' and  and ``The APC was funded by XXX''. Check carefully that the details given are accurate and use the standard spelling of funding agency names at \url{https://search.crossref.org/funding}, any errors may affect your future funding.}
%\dataavailability{\hl{Please add}}%In this section, please provide details regarding where data supporting reported results can be found, including links to publicly archived datasets analyzed or generated during the study. Please refer to suggested Data Availability Statements in section ``MDPI Research Data Policies'' at \url{https://www.mdpi.com/ethics}. You might choose to exclude this statement if the study did not report any data.} Not applicable.
\acknowledgments{NPN acknowledges the support of CNPq of Brazil under grant PQ-IB
309073/\mbox{2017-0.}}
\conflictsofinterest{The authors declare no conflict of interest.}%Declare conflicts of interest or state ``The authors declare no conflict of interest.'' Authors must identify and declare any personal circumstances or interest that may be perceived as inappropriately influencing the representation or interpretation of reported research results. Any role of the funders in the design of the study; in the collection, analyses or interpretation of data; in the writing of the manuscript, or in the decision to publish the results must be declared in this section. If there is no role, please state ``The funders had no role in the design of the study; in the collection, analyses, or interpretation of data; in the writing of the manuscript, or in the decision to publish the~results''.}

\end{paracol}
\reftitle{\hl{References}}%mdpi: please add title name for all the journal type refs

\end{document}